\documentclass[11pt,a4paper]{article}
\usepackage{prem}
\definecolor{kul-blue}{RGB}{29,141,176}

\title{Pricing service maintenance contracts using predictive analytics 
}

\author[1]{Laurens Deprez\footnote{corresponding author: \textit{laurens.deprez@kuleuven.be}}}
\author[1,2]{Katrien Antonio}
\author[1,3]{Robert Boute}
\affil[1]{\small{Faculty of Economics and Business, KU Leuven, Belgium.}}
\affil[2]{\small{Faculty of Economics and Business, University of Amsterdam, The Netherlands.}}
\affil[3]{\small{Technology \& Operations Management Area, Vlerick Business School, Belgium.}}
\date{}
\tikzstyle{decision} = [diamond, draw, fill=white, 
text width=4.5em, text badly centered, node distance=3cm, inner sep=0pt]
\tikzstyle{block} = [rectangle, draw, 
text width=5em, text centered, rounded corners, minimum height=3em]
\tikzstyle{block2} = [rectangle, draw, 
text width=7em, text centered, rounded corners, minimum height=3em]
\tikzstyle{block_small} = [rectangle, draw, 
text width=12em, text centered, rounded corners, minimum height=3em]
\tikzstyle{line} = [draw, -latex']
\tikzstyle{cloud} = [draw, ellipse,fill=red!20, node distance=3cm,
minimum height=2em]

\begin{document}
\maketitle

\begin{abstract}
\noindent 
As more manufacturers shift their focus from selling products to end solutions, full-service maintenance contracts gain traction in the business world. 
These contracts cover all maintenance related costs during a predetermined horizon in exchange for a fixed service fee and relieve customers from uncertain maintenance costs. 
To guarantee profitability, the service fees should at least cover the expected costs during the contract horizon.
As these expected costs may depend on several machine-dependent characteristics, e.g. operational environment, the service fees should also be differentiated based on these characteristics.
If not, customers that are less prone to high maintenance costs will not buy into or renege on the contract.
The latter can lead to adverse selection and leave the service provider with a maintenance-heavy portfolio, which may be detrimental to the profitability of the service contracts.
We contribute to the literature with a data-driven tariff plan based on the calibration of predictive models that take into account the different machine profiles. 
This conveys to the service provider which machine profiles should be attracted at which price. 
We demonstrate the advantage of a differentiated tariff plan and show how it better protects against adverse selection.
\iffalse
We study maintenance full-service agreements that cover all maintenance costs over a predetermined time horizon in exchange for a fixed premium. 
The stochastic and machine-dependent nature of the maintenance costs renders the determination of the premium challenging for the service provider. 
Inspired by insurance pricing, we use predictive analytics to determine the break-even price based on customer- and machine-dependent characteristics. 
We build a simulation engine to mimic the maintenance occurrences and their costs in a real environment and show the economic value of a differentiated pricing strategy.
\fi
\bigbreak
\noindent 
%\keywords{servitization, full-service contracts, contract pricing, predictive analytics, data-driven, risk management, after-sales support, maintenance-repair}
\end{abstract}

%A Due to the stochastic nature of the maintenance costs the determination of the correct break-even price of such a contract is a key challenge.  In this paper we present a methodology to price full-service (maintenance) contracts using predictive analytics to include customer characteristics. This approach is validated and the benefits of a price-differentiated strategy are shown using a simulated dataset.

\section{Introduction}
\label{sec: introduction}
%Full-service maintenance contracts are common practice in industry involving the maintenance of capital goods.
Product maintenance plays a central role in the business model of several manufacturing firms as they change their focus from selling products to selling product-service bundles. 
In the latter case, firms are responsible for the maintenance and functionality of their products.
This \emph{servitization} trend is motivated by the need to escape the product commoditization trap \citep{kastalli2013servitization}. 
Service providers and OEMs in the capital goods industry offer a range of service contracts to suit the needs of their customers.
Maintenance can either be provided by on-call service, which means that each preventive or corrective intervention is charged to the equipment user based on materials used and time spent. 
Alternatively, maintenance can be offered under a service plan such as a full-service contract.
A full-service contract covers all future costs of preventive and corrective maintenance, over a predetermined time horizon in exchange for a (series of) fee(s) or fixed premium(s).
Potentially, these contracts also include down-time compensation, to protect against the moral hazard of being served later than on-call customers.
In practice, a company could offer a combination of an on-call and a full-service contract.
This hybrid contract could for example only cover periodic maintenance or only certain failure types.
Examples of companies shifting their focus on services are General Electric Co., Siemens AG, and Hewlett-Packard Co. \citep{sawhney2003creating}.

In this paper, we study full-service contracts from the viewpoint of the service provider, who offers (sells) the contract to the user of the equipment. 
Under a full-service contract, the user no longer bears the risk of stochastic maintenance costs in exchange for the certitude of a fixed service fee; shifting the risk to the service provider. 
For the service provider, these contracts offer a guaranteed revenue stream.
They are also profitable in case the total premium volume of a portfolio, i.e.\ the collection of machines covered by a full-service contract, exceeds the maintenance costs.

Accurate price setting is a major challenge for service providers offering full-service maintenance contracts. 
Simply pricing full-service contracts based on historical estimation of expected total costs incurred over the contract period plus a safety margin (e.g. a factor reflecting the risk-averseness of the service provider times the standard deviation of the historical total costs) does not take the heterogeneity between machines and customers into account.
On the one hand, estimating expected costs for only a subset of machines with equal characteristics (by subdividing the data into different classes), stratifies the data into small sample sets for which costs cannot be accurately estimated \citep{denuit2007actuarial}. 
As a consequence, a justified price differentiation for a specific customer or machine is less straightforward. 
On the other hand, if all customers pay the same price for a contract, this may lead to adverse selection. 
\textquoteleft Low-risk\textquoteright\ customers, who require less maintenance and thus should be offered a lower price than maintenance-heavy customers, may not sign a full-service contract when they pay the same premium as the high-risk customers. As a result proportionally more high-risk customers will remain in the portfolio. 
This phenomenon is also observed in insurance and is detrimental for the profitability of a service contracts portfolio. 
Market research on services suggests that coming up with profitable service solutions is hard \citep{gebauer2005overcoming,hancock2005better,ulaga2011hybrid}; it seems that inadequate pricing makes it hard to exploit the financial potential of extended services \citep{rapaccini2015pricing}.

In an era of big data and data analytics the collection and statistical analysis of data can provide useful insights to many decision support systems, including tariff plans of full service contracts.
We propose a method to determine the price, or premium, of full-service contracts inspired by insights and techniques developed in the actuarial literature. 
A full-service maintenance contract can be considered an insurance covering the maintenance costs during a certain period of time.    
%The insurance literature therefore provides a rich source of statistical techniques which are inspiring for the case of pricing of full-service maintenance contracts. 
A key concept in insurance pricing that can be carried over to pricing full-service maintenance contracts, is the frequency-severity approach of handling claims or incurred costs \citep[see e.g.][]{henckaerts2018data,verbelen2018unraveling}. 
This involves modelling and predicting the frequency of failures and their associated costs, i.e. severity, separately and (statistically) independent.
%This frequency and severity approach can also be found in \citet{huber2012pricing} for repair contracts and in \citet{luo2017value} for warranty contracts.
A second element of insurance pricing is the use of \textquoteleft risk factors\textquoteright\ to reflect the heterogeneity of the risks in the portfolio \citep{henckaerts2018data}. 
To avoid lapses in a competitive market, insurance companies use rating factors (e.g. age, postal code area) to classify risks and to differentiate prices of their insurance products. 
Pricing through risk classification is the mechanism for insurance companies to compete as it allows an insurer to express which risk profiles should be attracted at which price.
Insurance companies therefore maintain large databases with policy(holder) characteristics, i.e.\ contract and customer information, and claim histories to build personalized risk-based pricing models. 
%Pricing is challenged by new evolutions in data availability and an increasing focus on individual risk based pricing. 

Using predictive analytics we develop an approach similar to insurance analytics that takes into account risk factors to predict the machine-specific costs under a full-service contract, and price it accordingly. 
For example, the reliability of a machine can be influenced by operating conditions, operator skills, service history and quality, among others. 
Some risk factors, such as temperature or humidity, can have a direct influence on the reliability.
Other risk factors are proxies for the real influencing factor. 
For instance, the machine location can be a proxy for the environmental conditions.
This information can be collected either by sensors installed on the machines, e.g. temperature measurements, or by sales people or technicians, at the time of sales or at a maintenance intervention respectively.
The latter is known as the Internet of people (IoP) and the former as the Internet of things (IoT).

This idea is new in the context of pricing maintenance contracts and will be used to detect which risk factors indicate significant differences in the expected number of maintenance interventions (frequency) and the expected costs incurred (severity). 
Including these risk factors allows a machine and customer specific tariff plan based on proper risk assessment. 
A differentiated tariff plan taking the heterogeneity of the machine portfolio into account also provides a better protection against adverse selection and makes the offering of the service provider more competitive.

Our predictive data-driven approach relies on the availability of historical data. 
We will demonstrate our methodology on simulated data.
The use of a simulation data provides a controlled  environment to  study  the  performance of our methodology in different scenarios prior to applying it to real data. 
We develop a simulation engine that is capable to simulate both (time-to-)failure data, and repair and maintenance costs in a variety of different machine environments such that it captures the heterogeneity in the contract data induced by the risk factors.
%We choose to simulate our data in the most granular way possible to maximize the applicability of our methodology.
The resulting data format is similar to what can be  available to a service provider with a maintenance contract portfolio.
The latter has been verified with an industrial partner.
We apply our predictive analytics models to our simulated data to determine the break-even price, which is the minimum premium required to have a profitable contract portfolio and we show the economic value of this pricing approach.

\section{Literature}
The literature on service maintenance contracts makes a distinction between resource-based  service and performance-based contracts. 
With resource-based service, maintenance is priced based on the time and materials spent; it can either be on-call or by use of a full-service maintenance plan.
With on-call service the customer contacts the service provider in case of failure and is charged for the corresponding maintenance. 
In contrast to full-service plans, the customer bears all the maintenance risk. 
We focus in this article on resource-based full-service contracts.
Performance-based contracts ensure the availability (or up-time) of the equipment, and their compensation is based on the availability of the underlying product \citep{kim2017reliability}. 
The literature comparing resource-based and performance-based contracts indicates that product reliability is generally higher under performance-based contracts \citep{Guajardo2012,bakshi2015signaling}. 
\citet{chan2018contracting} study sales and service records of medical equipment.
Their  dataset  consists  of  service  records  for  700  devices  over  a  four-year  period  including  the maintenance visits and their associated part costs and labor. 
They analyze the frequency of maintenance visits using a Poisson regression model and their associated part and labor costs using an exponential regression model.
They find that moving from on-call service to full-service plans reduces reliability and increases maintenance costs. 
Not only do operators reduce the level of their own care of the equipment under full-service contracts, the service provider also expends more resources to the equipment.

\citet{rapaccini2015pricing} presents a general overview of the literature on pricing service contracts. 
He distinguishes three pricing approaches: the price can be based on the predicted \emph{costs} under the contract; it can be based on the perceived \emph{value} of the service contract; or it can be benchmarked against the price of similar contracts offered by \emph{competitors}. 
Our method is a cost-based pricing approach, based on the expected costs of maintenance and repair, which can then be augmented by a desired risk and profit margin. 
The stochastic and machine-dependent nature of the frequency and severity of failures makes this a non-obvious exercise.
When this pricing scheme is then benchmarked with competition, small adaptations can be made for specific machine profiles to ensure that the \textquoteleft desired\textquoteright\ profiles are attracted to the portfolio.

\citet{huber2012pricing,huber2014pricing} describe a value-based pricing approach for full-service repair contracts, which do not include preventive maintenance.
They first develop a cost-based price, based on the failures and their associated costs, which are modelled independently by a non-homogeneous Poisson process and a distribution with finite support. %(the repair cost is bounded from below by the basic diagnostic cost, and from above by the replacement cost).
The value-based price is then obtained by the customer's choice (and utility) of on-call versus full-service, by means of a mean-variance utility function.
\citet{huber2014pricing} come up with a competition-based price by considering competition with other service providers. 
Whereas \citet{huber2012pricing} set out a stochastic model for the frequency of failures and associated costs to price full-service repair contracts, they do not calibrate their model on (historical) maintenance data, neither do they assess the future costs in a contract by a predictive model. 

The non-renewing free replacement warranty policy, where the manufacturer repairs or replaces a product at no cost, is similar to full-service contracts. 
It also involves an upfront payment of the warranty price, which is included in the sales price. 
\citet{luo2018mean,luo2017value} optimize the warranty policy for different product types jointly, by determining the optimal warranty price and the optimal length of the warranty period. 
The frequency of the incoming warranty claims and their associated costs are modelled as statistically independent.
Although the latter is similar to actuarial pricing and the work by \citet{huber2012pricing,huber2014pricing}, \citet{luo2018mean,luo2017value} do not include product characteristics (as risk factors) in their analysis, nor do they provide a data-driven approach to optimize the warranty policy.

The asset covered by a full-service maintenance contract is a repairable system.
The reliability literature discusses how to model time-to-failure for these repairable systems taking into account the effect of both preventive and corrective maintenance actions.
We refer to \citet{lindqvist2006statistical}, \citet{wu2010linear} and \citet{doyen2011modeling} for an introduction to these models.
We use their approach to model the frequency of failures for full-service maintenance contracts.

%nor do they look at covariates to integrate risk assessment.
Motivated by the current state-of-the art in insurance pricing \citep[see][for an overview]{denuit2007actuarial,Jong2008generalized}, our paper extends the literature on service contracts by introducing a data-driven pricing methodology to accommodate price differentiation based on a proper risk assessment using predictive analytics.
This is also in line with \citet{luo2017value}'s call for data-driven methodologies for warranty contracts and \citet{bertsimas2019predictive}'s insights on the inclusion of covariate data in problems in operations research and management science.

%On the one hand we consider the frequency of failure and the maintenance costs to be statistically independent, in line with \citet{huber2012pricing}, on the other hand our approach will also include predictive models. 
%These predictive models will be inspired by the regression models used in actuarial science and medical statistics for frequency and severity modelling.
%{\color{red} The rest of the paper has the following structure.}
\FloatBarrier
\section{Predictive analytics for full-service contracts}
\label{method}
In this section we describe the different aspects of a full-service maintenance contract, introduce its break-even price, and propose our methodology to calibrate this price using predictive models. 
Appendix \ref{appendix: notation} gives an overview of the key notations used in the paper.

\subsection{Problem description}
\label{stoch drivers}
We consider a service provider who provides maintenance to a collection of machines, potentially of different type, under a full-service contract.
We denote this collection of full-service contracts as the contract portfolio.
A full-service contract covers all maintenance and repair costs for that machine during a period $[t_0,t_0+\Delta t]$, with $\Delta t$ the finite and predetermined duration of the contract.
The start of the contract $t_0$ can either be the moment of the machine sale or a later moment.
The start $t_0$ can be different for the different machines in the portfolio.
During this coverage period $\Delta t$, $n_m(\Delta t)$ preventive maintenance interventions are included in the contract, which are typically also scheduled at initiation of the contract. 
Whereas $n_{m}(\Delta t)$ is usually deterministic, the associated costs of preventive maintenance, $S_{j}$ for $j \in \{1,2,..., n_m(\Delta t)\}$, are not.
We will assume that each preventive maintenance is of the same type and has the same effectiveness.
This does not mean that all preventive maintenances have exactly the same cost -- we assume they are intrinsically stochastic, but this difference in costs is not related to the effectiveness; as a result in our model the cost of the preventive maintenance does not impact the subsequent failure intensity.
We note that the influence of preventive maintenance actions could be further incorporated via time-varying covariates in our statistical models. 
These time-varying covariates could represent e.g.\ the time since last maintenance event, the number of maintenance events in the past year, the cost of the last preventive maintenance, etc.

The contract additionally covers the costs of the corrective maintenance due to unplanned failures and/or due to malfunctioning of a component of the machine that requires repair or even replacement. 
These can be minor failures, requiring minimal corrective maintenance, or catastrophic failures requiring a major overhaul of the machine. 
Both the corrective maintenance and the overhaul are unplanned and consequently their occurrence and timing are stochastic.
We distinguish between $n_f$ different types of failures, and denote $N_{f,i}(\Delta t)$ the number of failures of type $i$ that occur during the contract period $[t_0,t_0+\Delta t]$ and $X_{i,k}$ the associated cost of the $k$th failure of type $i$. 
Both $N_{f,i}(\Delta t)$ and $X_{i,k}$ are unknown at initiation of the contract.

\begin{figure}[tb]
	\centering
	\begin{tikzpicture}[scale=0.8,snake=zigzag, line before snake = 5mm, line after snake = 5mm]
	% axis and titles
	\draw [->] (0,0)--(12,0);
	\node [below] at (0,-.1) {$t$};
	\node [below] at (11,-.1) {$t+\Delta t$};
	\node [] at (11,0.4) {time};
	\draw [-] (0,0)--(0,-.1);
	\draw [-] (11,0.1)--(11,-.1);
	\draw [-] (-.1,0)--(0,0);
	\node [left] at (-.2,0) {$0$};
	\draw [->] (0,0)--(0,6);
	\node [rotate=90,right] at (0.4,5) {costs};
	% maintenance
	\draw [black,thick] (2,0)--(2,2.4);
	\draw [black,thick] (4,0)--(4,2.85);
	\draw [black,thick] (6,0)--(6,2.05);
	\draw [black,thick] (8,0)--(8,2.6);
	\node [black] at (2,2.4) {\tiny{$\bullet$}};
	\node [black] at (4,2.85) {\tiny{$\bullet$}};
	\node [black] at (6,2.05) {\tiny{$\bullet$}};
	\node [black] at (8,2.6) {\tiny{$\bullet$}};
	\draw [dashed,line width=0.1mm,gray] (0,2.4)--(2,2.4);
	\draw [dashed,line width=0.1mm,gray] (0,2.85)--(4,2.85);
	\draw [dashed,line width=0.1mm,gray] (0,2.05)--(6,2.05);
	\draw [dashed,line width=0.1mm,gray] (0,2.6)--(8,2.6);
	\node [black,below,font=\sffamily] at (2,-.1) {$m_1$};
	\node [black,below] at (4,-.1) {$m_2$};
	\node [black,below] at (6,-.1) {$m_3$};
	\node [black,below] at (8,-.1) {$m_4$};
	\node [black] at (2,0) {\tiny{$\bullet$}};
	\node [black] at (4,0) {\tiny{$\bullet$}};
	\node [black] at (6,0) {\tiny{$\bullet$}};
	\node [black] at (8,0) {\tiny{$\bullet$}};
	\node [black] at (0,2.4) {\tiny{$\bullet$}};
	\node [black] at (0,2.85) {\tiny{$\bullet$}};
	\node [black] at (0,2.05) {\tiny{$\bullet$}};
	\node [black] at (0,2.6) {\tiny{$\bullet$}};
	\node [black,right] at (2,2.3) {$S_1$};
	\node [black,right] at (4,2.85) {$S_2$};
	\node [black,right] at (6,2.05) {$S_3$};
	\node [black,right] at (8,2.6) {$S_4$};
	% failure type 1
	\draw [red,thick] (3,0)--(3,3.7);
	\draw [red,thick] (9,0)--(9,3.1);
	\node [red] at (3,3.7) {\tiny{$\bullet$}};
	\node [red] at (9,3.1) {\tiny{$\bullet$}};
	\node [red,above] at (3,3.7) {$X_{1,1}$};
	\node [red,above] at (9,3.1) {$X_{1,2}$};
	\node [red,below,font=\sffamily] at (3,-.1) {$f_{1,1}$};
	\node [red,below] at (9,-.1) {$f_{1,2}$};
	\node [red] at (3,0) {\tiny{$\bullet$}};
	\node [red] at (9,0) {\tiny{$\bullet$}};
	\draw [dashed,line width=0.1mm,gray] (0,3.7)--(3,3.7);
	\draw [dashed,line width=0.1mm,gray] (0,3.1)--(9,3.1);
	\node [red] at (0,3.1) {\tiny{$\bullet$}};
	\node [red] at (0,3.7) {\tiny{$\bullet$}};
	% failure 2
	\draw [magenta,thick] (7,0)--(7,4.8);
	\draw [dashed,line width=0.1mm,gray] (0,4.8)--(7,4.8);
	\node [magenta,below] at (7,-.1) {$f_{2,1}$};
	\node [magenta,above] at (7,4.8) {$X_{2,1}$};
	\node [magenta] at (7,4.8) {\tiny{$\bullet$}};
	\node [magenta] at (0,4.8) {\tiny{$\bullet$}};
	\node [magenta] at (7,0) {\tiny{$\bullet$}};
	% define and draw distributions
	\def\normalone{\x,{.5*1/exp(((\x-2)^2)/.5)}}
	\def\normaltwo{\x,{.5*1/exp(((\x-3)^2)/.5)}}
	\def\normalPM{\x,{.5*1/exp(((\x-2.5)^2)/.5)}}
	\draw[domain=1.501:5,smooth,variable=\x,black,rotate=90,very thick] plot ({\x},{2*1/(\x-1.5)/sqrt(2*pi)/1*exp(-(ln(\x-1.5))^2/2/1^2)});
	\draw[domain=2.501:6.5,smooth,variable=\x,red,rotate=90,very thick] plot ({\x},{3*1/(\x-2.5)/sqrt(2*pi)/.8*exp(-(ln(\x-2.5))^2/2/.8^2)});
	\draw[domain=3.501:6.5,smooth,variable=\x,magenta,rotate=90,very thick] plot ({\x},{2*1/(\x-3.5)/sqrt(2*pi)/.8*exp(-(ln(\x-3.5))^2/2/.8^2)});
\end{tikzpicture}
	\caption{A full-service contract covers all costs of preventive maintenance interventions $m_j$ and unplanned failures $f_{i,k}$, of failure type $i$, with respective costs $S_j$ and $X_{i,k}$ during the coverage period $[t_0,t_0+\Delta t]$.}
	\label{fig: machine life}
\end{figure}
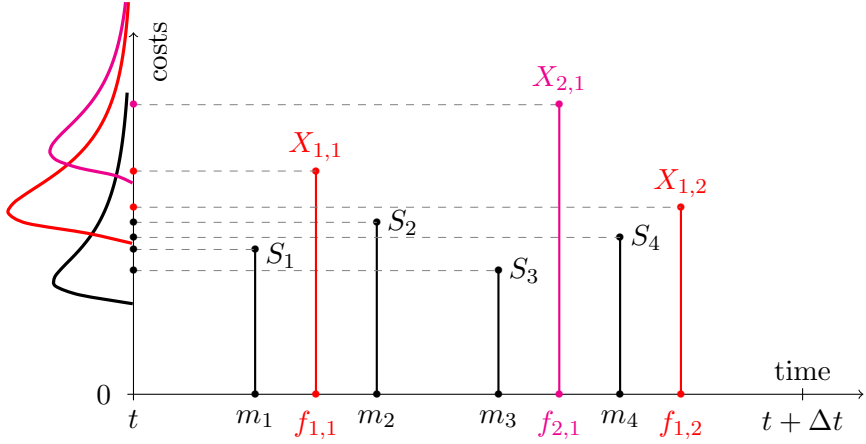

Figure~\ref{fig: machine life} illustrates the timeline of a contract during the coverage period $[t_0,t_0+\Delta t]$, with on the horizontal axis (indicating the time dimension) the planned preventive maintenance interventions $m_j$ and the unplanned failures of type $i$, $f_{i,k}$.
The vertical lines, for every event, represent the size of the associated costs, $S_j$ and $X_{i,k}$ respectively.
The probability density functions on the vertical axis illustrate the stochastic nature of these costs.

The total costs covered by the service contract, $C(\Delta t)$, equal the total aggregated failure costs $F(\Delta t)$  and the total aggregated preventive maintenance costs, $M(\Delta t)$, covered by the contract over the period $\Delta t$:
\begin{equation}
C(\Delta t) = F(\Delta t)+M(\Delta t),
\label{eq: total costs}
\end{equation}
with  
\begin{equation}
F(\Delta t) = \sum_{i=1}^{n_{f}}\sum_{k=1}^{N_{f,i}(\Delta t)}X_{i,k},
\end{equation}
and
\begin{equation}
M(\Delta t) = \sum_{j=1}^{n_{m}(\Delta t)}S_{j}.
\end{equation}
We consider the occurrence of a failure and its associated cost to be statistically independent, in line with most literature \citep{huber2012pricing,luo2018mean,luo2017value,henckaerts2018data}.
Taking the dependence between frequency and severity into account is much more involved. This has for instance been done by \citet{gschlossl2007spatial} and \citet{czado2012mixed}.
The assumption of independence between severity and frequency of failures is typical in insurance pricing and makes the frequency and severity analysis more tractable.

The preventive maintenance scheme influences the occurrence of failures and consequently affects the failure costs $F(\Delta t)$.
We consider this maintenance scheme exogenous \citep[as in e.g.][]{poppe2018hybrid} and focus on determining the premium given this maintenance scheme.
When this preventive maintenance scheme would be optimized by trading off the cost of preventive maintenance with the cost of failure \citep[see e.g.][]{barlow1960,rebaiaia2017periodic}, the expected costs and resulting break-even price are adapted accordingly.

%the price of the maintenance model although it is not our objective to alter nor optimize this scheme.

\begin{figure}[tb]
    \centering
    \begin{tikzpicture}
	\definecolor{lightb}{RGB}{153,204,255}
	\definecolor{dm}{RGB}{222, 89, 113}
	\definecolor{m}{RGB}{255,0,255}
	\draw [fill = black, black] (0,-0.05) rectangle (0.1,0.05) {};
	\node [right] at (0.1,0.04) {\small{maintenance}};
	\draw [fill = red, red] (2.3,-0.05) rectangle (2.4,0.05) {};
	\node [right] at (2.4,0.01) {\small{failure type 1}};
	\draw [fill = m, m] (4.8,-0.05) rectangle (4.9,0.05) {};
	\node [right] at (4.9,0.01) {\small{failure type 2}};
%	\draw [fill = dm, dm] (7.3,-0.05) rectangle (7.4,0.05) {};
%	\node [right] at (7.4,0.01) {\small{failure type 3}};
	\end{tikzpicture}
    \includegraphics[width =\textwidth]{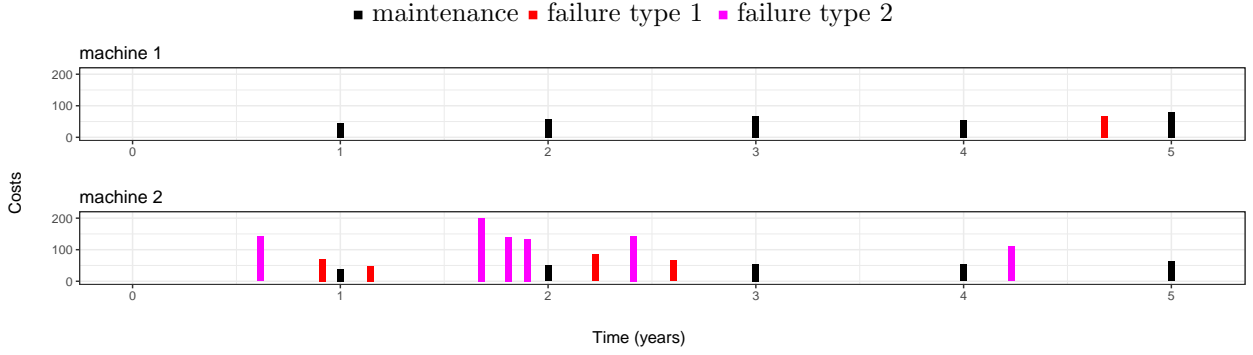}
    \caption{Two simulated timelines, where the heterogeneity in the machine population is reflected by different maintenance costs during the contract horizon of 5 years.}
    \label{fig: heterogeneous}
\end{figure}

%\subsubsection*{Heterogeneous population} \label{heterogeneity}
The heterogeneity between machines in the portfolio is reflected in the variety in costs, $C(\Delta t)$ incurred for different machines.
Some machines may incur more failures and/or costs than others, as Figure~\ref{fig: heterogeneous} illustrates. 
In Figure~\ref{fig: heterogeneous}, machine 2 is more expensive to maintain during the contract horizon than machine 1 due to more failures.
This heterogeneity in the costs motivates the need for a differentiated tariff plan.
We will use observable and measurable risk factors to reflect this heterogeneity and set prices accordingly.
These risk factors can be both static, e.g. the country of residence, operational environment, operator and maintenance crew skill level, as well as dynamic over time, e.g. the temperature of the machine, maintenance and service history \citep{barabadi2012reliability,barabadi2014application,nouri2016tire}.
This requires the service provider to keep track of the values of these risk factors for each machine, at contract initiation (static) as well as over time (dynamic). 

\FloatBarrier
\subsection{Technical price of full-service contracts}
\label{pricing scheme}
Denote $P(\Delta t)$ the premium paid at initiation of the contract, and $R(\Delta t)$ the profit at termination of the contract,
\begin{equation}
R(\Delta t) = P(\Delta t)-C(\Delta t).
\end{equation}
The break-even or technical price $P^\star(\Delta t)$ is the premium for which the expected profit $E[R(\Delta t)] = 0$.
The commercial price $P(\Delta t)$ includes additional loadings added to the break-even price $P^\star(\Delta t)$, e.g. risk margins, profit margins, commissions and expenses, adjustments based on a competitive analysis in collaboration with the marketing department.
Risk margins that account for the risk-averseness of the service provider and distributional properties of the incurred costs, can be determined by making use of so-called \textquoteleft premium principles\textquoteright\ \citep{kaas2008modern} studied in actuarial science.
The break-even price $P^\star(\Delta t)$ is expressed as
\begin{equation}
\begin{aligned}
P^\star(\Delta t) &= E[C(\Delta t)]= E\left[\sum_{i=1}^{n_{f}}\sum_{k=1}^{N_{f,i}(\Delta t)}X_{i,k} + \sum_{j=1}^{n_{m}(\Delta t)}S_{j}\right]\\
&= \sum_{i=1}^{n_{f}}E[N_{f,i}(\Delta t)]\cdot E[X_{i,k}] + n_{m}(\Delta t)\cdot E[S_j],
\end{aligned}
\label{eq: breakeven P}
\end{equation}
where the last equality assumes independence between the frequency of failure and their associated costs.
Equation \eqref{eq: breakeven P} shows how the break-even price depends on the expected frequency of failure $E[N_{f,i}(\Delta t)]$ and the number of preventive maintenance actions $n_m(\Delta t)$ on the one hand and the expected severity of failure and preventive maintenance, $E[X_{i,k}]$ and $E[S_j]$ resp., on the other hand. 
To account for machine heterogeneity, we need to determine the technical price $P^\star(\Delta t)$ conditional on the machine characteristics, which are represented by covariates that can be fixed, $\bm{x}_1$, or time-varying, $\bm{x}_2(t)$:
\begin{equation}
P^\star(\Delta t)|\bm{x}_1,\bm{x}_2(t) = \sum_{i=1}^{n_{f}}E[N_{f,i}(\Delta t)|\bm{x}_1,\bm{x}_2(t)]\cdot E[X_{i,k}| \bm{x}_1,\bm{x}_2(t)] 
+ n_{m}(\Delta t)\cdot E[S_{j}| \bm{x}_1,\bm{x}_2(t)].
\label{eq: cond break even price}
\end{equation}
In the above expression, the machine characteristics impact both the expected frequency of failure and the expected costs of maintenance and failure.
The resulting conditional break-even price $P^\star(\Delta t)|\bm{x}_1,\bm{x}_2(t)$ will then be different for each machine profile, reflecting the heterogeneity in the machine population.
%is reflected in the price list \colr{(what is price list?)}\colg{(de lijst van prijzen)}.

\iffalse
\begin{figure}[tb]
	\centering
	\input{tikz_workflow}
	\caption{We first build a simulation engine to generate failure and cost data for a variety of machine profiles (risk factors). We then calibrate predictive models for the frequency and severity of maintenance in the presence of these risk factors. The calibrated predictive models are used to determine the break-even price of the contract. Finally, we evaluate the economic benefit of a differentiated pricing scheme.\colr{do we update this or remove it!?}}
	\label{fig: workflow}
\end{figure}
\fi

\subsection{Data description}\label{sec: contract data}
To apply our pricing methodology, a dataset is required with a history of failures and preventive maintenance interventions, together with their costs.
%This will be used to calibrate predictive models in order to price (future) contracts.
We rely on maximum likelihood estimation to fit the time-to-failure, failure types and costs distribution on this dataset.
The purpose of this calibration is to obtain data driven values for the parameters used in the distributions, instead of fixing parameter values upfront without taking the available data into account.
This dataset stores the maintenance activities performed on each machine in the portfolio of, say $n$, machines, during an observation window $t_{\text{obs}}$.
Each machine has a number of characteristics or covariates $\bm{x}\ (= \bm{x}_{1} \cup \bm{x}_{2}(t))$, where we make a distinction between time-independent $\bm{x}_{1}$ and time-dependent $\bm{x}_{2}(t)$ covariates.
We consider $n_f = 1 + n_{f,\text{minor}}$ different failure types, consisting of $n_{f,\text{minor}}$ minor failure types on the one hand and a catastrophic failure on the other hand. 
The minor failures types, denoted $\{f_1,f_2,...,f_{n_{f,\text{minor}}}\}$, require minimal, as-good-as-old corrective maintenance \citep[see e.g.][]{lindqvist2006statistical,wu2010linear,doyen2011modeling}, which does not impact the overall reliability of the machine.
Catastrophic failures, denoted $f_c$, require a full overhaul of the machine and bring the machine in a new state.
The overhaul is considered a perfect corrective maintenance and the virtual age \citep{kijima1989some} of the machine is zero afterwards \citep{sheu2012generalised}.
Furthermore, the periodic maintenance events, denoted $m$, follow a time-based preventive maintenance strategy specified by the maintenance interval $\Delta t_{\text{PM}}$.
This fixes the number of periodic maintenance actions $n_m(\Delta t)$ during the contract duration $\Delta t$.
Table~\ref{tab: sim data} provides an excerpt of such a dataset.
Each line corresponds to a failure or preventive maintenance and takes the form: $(i,t,t_{-},c,y,X,$ $\bm{x}_{1}, \bm{x}_{2}(t))$, with $i$ the machine number, $t$ the event time, $t_{-}$ the time of the previous failure or maintenance (if the current event is respectively a failure or maintenance), $c$ the censoring status with $c=1$ if the event is observed and $c=0$ if the timeline is censored, $y\ \in \{m,f_1, f_2, ...,f_{n_{f,\text{minor}}},f_c\}$ the type of the event, $X$ the costs, and $\bm{x}_{1}$ and $\bm{x}_{2}(t)$ the values of respectively the fixed and time-dependent covariates of a machine at the event time $t$. 
The timeline of a machine is censored at the end of a contract as well as at the end of the period of data collection, since both the ending of the contract as well as the historical data collection prevent the observation of the next failure.
In Table~\ref{tab: sim data}, the covariate $\bm{x}_{1}$ represents the (static) country code and $\bm{x}_{2}(t)$ indicates 1 if a type 3 failure has occurred in the past, and zero otherwise.
The event timeline of a single machine is defined by the set of all events registered on the same machine in the dataset, see Table \ref{tab: sim data}.
\begin{table}[tb]
\caption{Excerpt of a dataset required to price full-service contracts, detailing for each failure and preventive maintenance the machine number~$i$, the time stamp~$t$ of the event time and its previous occurrence~$t_{-}$, its censoring status~$c$, type~$y$ and costs~$X$, together with the values of the machine-dependent covariates, $\bm{x}_1$ and $\bm{x}_2(t)$.}
\label{tab: sim data}
	\centering
	\begin{tabular}{cccccccc}
		\toprule
		$i$ & $t$ & $t_{-}$ & $c$ & $y$ & $X$ & $x_{1}$ & $x_{2}(t)$ \\
		\midrule
		1 & 0.12 & 0 & 1 & $f_3$ & 229.31 & BE & 0 \\
		
		1 & 1 & 0 & 1 & $m$ & 63.06 & BE& 1\\
		
		1 & 1.53 & 0.12 & 1 & $f_2$ & 230.97 & BE &  1\\
		
		$\vdots$ &  & &  $\vdots$&  &  &  & $\vdots$ \\
		%1 & 2 & 1 & 1 & m & 54.08 & BE & 1\\
		
		%1 & 2.75 & 1.53 & 1 & 2 & 216.15 & BE & 1 \\
		
		%1 & 3 & 2 & 1 & m & 64.99 & BE & 1\\
		
		%1 & 3.09 & 2.75 & 1 & 2 & 141.64 & BE & 1\\
		
		1 &3.14 &3.09 & 0 &0 &0 & BE &1\\
		\midrule
		$\vdots$ &  & & $\vdots$ &  &  &  & $\vdots$ \\
		\midrule
		$n$ & 1 & 0 & 1 & $m$ & 46.98 & NL & 0\\
		
		$n$ & 2 & 1 & 1 & $m$ & 77.31 & NL & 0\\
		
		$n$ & 3 & 2 & 1 & $m$ & 54.10 & NL & 0\\
		
		$n$ & 3.49 & 0 & 1 & $f_3$ & 195.78 & NL & 0\\
		
		$\vdots$ &  & & $\vdots$ &  &  &  & $\vdots$ \\
		
		$n$ & 5 & 4.99 & 0 & 0 & 0 & NL & 1\\
		\bottomrule
	\end{tabular}
\end{table}
\FloatBarrier
\section{Calibration of predictive models for pricing}
\label{sec: calibration}
Making use of maximum likelihood estimation, we calibrate predictive models to the contract data, as described in Section \ref{sec: contract data}, capturing the failure occurrences, failure types, and the failure and maintenance costs observed over the machine-specific timelines. 
We hereby identify the impact of the machine-dependent characteristics $\bm{x}_1$ and $\bm{x}_2(t)$ to determine a machine-dependent technical (break-even) price $P^\star|\bm{x}_1,\bm{x}_2(t)$.

\subsection{Likelihoods of the predictive models}
We propose a hierarchical approach inspired by \citet{frees2008hierarchical} and decompose the joint probability distribution of the contract data in its stochastic elements: failure times and types, costs of failures and maintenance events. 
The joint probability distribution of the timeline of a single machine with failure times $\bm{T}$, failure types $\bm{Y}$, failure costs $\bm{X}$ and maintenance costs $\bm{S}$ denoted by $f(\boldsymbol{T},\boldsymbol{Y},\boldsymbol{X},\boldsymbol{S})$, can be decomposed as follows:
\begin{equation}
	f(\boldsymbol{T},\boldsymbol{Y},\boldsymbol{X},\boldsymbol{S})= f(\boldsymbol{T},\bm{Y})\cdot f(\boldsymbol{X}\vert \boldsymbol{T},\boldsymbol{Y})\cdot f(\boldsymbol{S}\vert \boldsymbol{T},\boldsymbol{Y},\bm{X}),
\end{equation}
where covariates are omitted to promote legibility.
%Alternatively, a contract time-line can be interpreted in a more granular way by considering the total number of failures $N$ during the contract duration $\Delta t$ instead of accounting for all specific failure times.
%In this case the total probability distribution takes the form $f(N, \Delta t, \boldsymbol{Y},\boldsymbol{X},\boldsymbol{S})$ and can be decomposed in similar fashion as
%\begin{equation}
%	f(N, e, \boldsymbol{Y},\boldsymbol{X},\boldsymbol{S})= f(N\vert \Delta t) \cdot f(\boldsymbol{Y}\vert N)  \cdot f(\boldsymbol{X}\vert \boldsymbol{T},\boldsymbol{Y})\cdot f(\boldsymbol{S}).
%\end{equation}
%Although this approach takes into account less detailed, it is more straightforward to implement.

\paragraph{Time-to-failure likelihood}
The first component $f(\bm{T},\bm{Y})$ deals with the failures times and types.
First, we analyze the failure times while only distinguishing between minor, denoted $f_m$, and catastrophic, denoted $f_c$, failures.
To price full-service maintenance contracts, it is not necessary to know the exact failure times, it suffices to estimate the number of failures during the contract period, see Equation \eqref{eq: breakeven P}.
This gives us two possible approaches to model the data on occurrences of failures as observed over the lifetime of a machine or service contract. 
On the one hand we can propose and calibrate a time-to-failure model which uses the full detail of the failure times, as displayed in Table \ref{tab: sim data}.
Such a model does not only allow to assess the number of failures during the contract horizon, but also to predict the failure times.
On the other hand, aggregating the time-to-failure data into discrete time periods allows the use of count models. 
Such regression models predict the number of failures in a next time period.
In line with the granularity of the data in Table \ref{tab: sim data}, we opt for time-to-failure models as they provide more detail than count models. 

To calibrate the time-to-failure model, we maximize its specific time-to-event log-likelihood.
Each failure in the dataset is characterized by a vector $(y,t,t_{0},c)$, where $y\in \{f_m,f_c\}$ is the failure type, $t$ indicates the failure time, $t_{0}$ the time of the previous failure and $c \in \{0,1\}$, with $0$ indicating censoring of the event.
In first instance we aggregate the different minor failure types, $\{f_1,f_2,...,f_{n_{f,\text{minor}}}\}$, as a single type $f_m$.
The contribution of each failure to the log-likelihood is,  
\begin{equation}
c\log(\lambda_y(t)) + \log(S_{T}(t\vert t_{0})).
\label{eq: lik}
\end{equation}
Summing over all machines and failures gives the full log-likelihood.
$\lambda_y(t)$ is the type-specific failure intensity function and
$S_{T}(t\vert t_0)$ is the survival function for the time-to-next-failure,
\begin{equation}
    S_{T}(t\vert t_0) = \exp\left(-\int_{t_0}^t\lambda_{\text{total}}(u)du\right).
\end{equation}
The total failure intensity $\lambda_{\text{total}}(t)$ is the sum of the type-specific failure intensity functions for the minor and the catastrophic failures and totally specifies the survival function $S_T(t\vert t_0)$ for the time-to-next-failure.

\begin{figure}[tb]
	\centering
	\includegraphics[width = \textwidth]{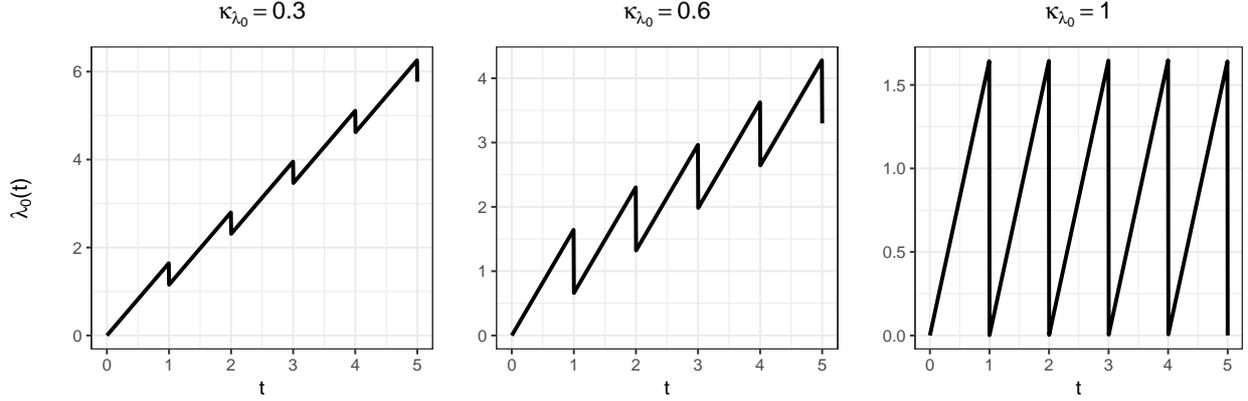}
	\caption{Baseline failure intensity $\lambda_0(t)$ for scale $\alpha_{\lambda_0} = \exp(0.5)$, intercept $\gamma_{\lambda_0} = 0$, maintenance interval $\Delta t_\text{PM} = 1$ and different values of $\kappa_{\lambda_0}\ (\in \{0.3,0.6,1\})$.}
	\label{fig: hazards}
\end{figure}

The type-specific failure intensity function $\lambda_{y}(t)$ for the minor failures, denoted $\lambda_{f_m}(t)$, is machine-specific and depends on the machine characteristics $\bm{\chi}_{1} \subseteq \bm{x}_{1}$ and $\bm{\chi}_{2}(t)\subseteq \bm{x}_{2}(t)$ as follows:
\begin{equation}
	\lambda_{f_m}(t) = \lambda_0(t)\exp(\bm{\beta}'_1\cdot \bm{\chi}_{1}+\bm{\beta}'_2\cdot \bm{\chi}_{2}(t)).
	\label{eq: minor failure hazard}
\end{equation}
We denote $\bm{\chi}_{1}$ and $\bm{\chi}_{2}(t)$ as the subset of the machine characteristics incorporated in the failure intensity for minor failures.
Not all characteristics available in the data necessarily influence the failure intensity.
This failure intensity function consists of a baseline $\lambda_0(t)$, i.e. common to all machines, and proportional influence of the covariates \citep{cox1975partial}.
The effect of the covariates scales the failure intensity function but does not influence its functional form.
The baseline failure intensity $\lambda_0(t)$ depends on the preventive maintenance scheme, defined by the maintenance interval $\Delta t_{\text{PM}}$. 
For a time-based imperfect preventive maintenance with interval $\Delta t_\text{PM}$, the baseline failure intensity is characterized by \citep{lindqvist2006statistical,wu2010linear}:
\begin{equation}
\lambda_0(t) = \alpha_{\lambda_0}\cdot\left((t\ \text{mod}\ \Delta t_\text{PM}) + (1-\kappa_{\lambda_0})
\left\lfloor \frac{t}{\Delta t_\text{PM}}\right\rfloor\right) + \gamma_{\lambda_0},
\label{eq: imperfect hazard}
\end{equation}
where  $\alpha_{\lambda_0} \in \mathbb{R}^+_0$ is the scale, $\gamma_{\lambda_0}\in \mathbb{R}^+$ is the intercept and $\kappa_{\lambda_0}\in [0,1]$ is the improvement after each maintenance.
The modulo operator mod returns the remainder after Euclidian division.
The floor operator $\lfloor \cdot \rfloor$ gives as output the greatest integer smaller than the argument.
Figure \ref{fig: hazards} illustrates the influence of $\kappa_{\lambda_0}$ on the baseline failure intensity function.
After each preventive maintenance action, the failure intensity decreases with $\kappa_{\lambda_0}$ percentage.
Consequently, the effectiveness of each preventive maintenance action is the same.
The closer $\kappa_{\lambda_0}$ gets to one, the better the maintenance;
if $\kappa_{\lambda_0}=1$ then the maintenance is perfect.

The type-specific failure intensity function $\lambda_y(t)$ for the catastrophic failures, denoted $\lambda_{f_c}(t)$, is modelled as follows
\begin{equation}
	\lambda_{f_c}(t) = \kappa_{c}\alpha_{c}^{\kappa_{c}}t^{\kappa_{c}-1}.
	\label{eq: weibull hazard}
\end{equation}
Here we assume a Weibull failure intensity, which is widely used in reliability engineering due to its versatility and relative simplicity \citep{bobbio1980modeling,wang2000application,wu2019failure}, but other choices can be valid. 
In our model the risk factors do not influence the catastrophic failure mode, however adding this influence is a possible extension of this model.

Maximizing the log-likelihood gives estimates for the model parameters, i.e. $\alpha_{\lambda_0},\kappa_{\lambda_0}$ and $\gamma_{\lambda_0}$ for the baseline hazard, $\bm{\beta}_1$ and $\bm{\beta}_2$ for the influence of the machine covariates, and $\alpha_{c}$ and $\kappa_{c}$ for the catastrophic failures.

\paragraph{Likelihood for minor failure types}
The component $f(\bm{T},\bm{Y})$ also contains the failure types.
Previously, we incorporated only the distinction between minor and catastrophic failures.
Now we handle the different minor failure types.
We estimate the parameters of a multinomial logit model which delivers the probabilities for the minor failure types $Y\vert (Y\in \mathcal{N}_{f,\text{minor}})$, where $\mathcal{N}_{f,\text{minor}} = \{f_1,f_2,...,f_{n_{f,\text{minor}}}\}$ \citep{frees2008hierarchical}.
Using the observed data, the contribution to the log-likelihood of an observed minor failure type $y$ is 
\begin{equation}
    \log \left(\Pr[Y = y\vert Y\in \mathcal{N}_{f,\text{minor}}]\right).
	\label{eq: logit lik}
\end{equation} 
Summing over all machines and minor failures gives the full log-likelihood.
Following \citet{frees2008hierarchical}, these probabilities are specified as
\begin{equation}
\begin{aligned}
\Pr[Y = y\vert Y \in \mathcal{N}_{f,\text{minor}},\bm{z}]  &= \frac{\exp( \boldsymbol{\alpha}_y^{'} \boldsymbol{z}) }{1 + \sum_{n = 1}^{n_{f,\text{minor}}-1} \exp( \boldsymbol{\alpha}_n^{'} \boldsymbol{z})}, \ \forall y < n_{f,\text{minor}}\\
\Pr[Y = n_{f,\text{minor}}\vert Y \in \mathcal{N}_{f,\text{minor}}, \bm{z}]  &= \frac{1}{1 + \sum_{n = 1}^{n_{f,\text{minor}}-1} \exp( \boldsymbol{\alpha}_n^{'} \boldsymbol{z})},
\end{aligned}
\label{eq: multinom prob}
\end{equation}
where $\bm{z}$ denotes the covariates influencing the failure type probabilities.
$\bm{z}$ can include a selection of both fixed and time-dependent covariates, 
where the time-dependent covariates are evaluated at the failure time.
These probabilities specify a multinomial distribution which is typically used to classify multiclass data.
If multiple minor failures occur at the same time, they are modelled as a single minor (extra) failure type, e.g. type 3 could be the co-occurrence of a type 1 and a type 2 failure.
Maximizing the log-likelihood gives estimates for the parameters $\bm{\alpha}_y$ driving each minor failure type $y\ (\in \mathcal{N}_{f,\text{minor}})$.\footnote{In the R \citep{Rsoftware} package \textquoteleft nnet\textquoteright\ \citep{nnet}, maximum likelihood estimation for the multinomial logit model is available.}

\paragraph{Severity likelihood}
The last two components of the decomposition $f(\boldsymbol{X}\vert \boldsymbol{T},\boldsymbol{Y})$ and \\
$f(\boldsymbol{S}\vert\boldsymbol{T},\boldsymbol{Y}, \boldsymbol{X})$ deal with the failure and maintenance costs respectively.
We compose a distributional model for the costs using marginal severity distributions on the one hand and a copula for the dependency of costs in case of co-occurring failures on the other hand. 
\iffalse
The contribution of each observed preventive maintenance cost $s$ to the log-likelihood is,
\begin{equation}
	\log\left(f_S(s)\right),
	\label{eq costs lik}
\end{equation}
where $f_S(s)$ is the probability density function associated with the maintenance costs.
Summing over all machines and preventive maintenance events gives the full log-likelihood.
\fi
We model the probability density function of the preventive maintenance costs with a positive and right-skewed distribution, e.g.\ a gamma distribution,
\begin{equation}
\begin{aligned}
S &\stackrel{d}{\sim} f_S(\mu,\sigma) \\
%X_{y,k} &\stackrel{d}{\sim} f_{X_{y}}(\mu_y,\sigma_y),\ y \in \mathcal{N}_{f,\text{minor}}\\
%X_{y,k} &\stackrel{d}{\sim} f_{X_y}(\mu_c,\sigma_c),\ y = \text{catastrophic},
\end{aligned}
\label{eq: costs distr}
\end{equation}
with $\mu$ the location parameter and $\sigma$ the scale parameter.
We use the following parameterization for the location parameters: $g(\mu) = \boldsymbol{\gamma}^{'}\boldsymbol{w}$, where the link function $g(.)$ is chosen in such a way that the location is positive.
We include heterogeneity in these costs through a set of contract specific covariates $\boldsymbol{w}$.
Once again, these covariates $\bm{w}$ can include a selection of both fixed and time-dependent covariates, 
where the time-dependent covariates are evaluated at the failure time.
Maximizing the log-likelihood gives estimates for $\bm{\gamma}$.
A similar approach can be followed for minor failure costs $X_{y}$ (with $y \in \mathcal{N}_{f,\text{minor}})$ and the catastrophic failure costs $X_{y}$ (with $y = f_c)$.
We make a distinction between the cost distribution of the minor and catastrophic failures, because they are different in nature.
We denote the influencing covariates on the minor and catastrophic failures as $\bm{w}_y$ and $\bm{w}_c$ respectively.
The parameters of the cost distributions of the minor and catastrophic failures are denoted as $\mu_y$, $\sigma_y$ and $\gamma_y$ (with $y \in \mathcal{N}_{f,\text{minor}})$ and $\mu_c$, $\sigma_c$ and $\gamma_c$ respectively.
A copula with positive dependence, e.g.\ the Frank copula, captures the statistical dependence between the multiple minor failures costs occurring at the same time \citep{frees2008hierarchical}.
We opt for a positive dependence since we assume high failure costs in a failure type will induce high failure costs in the other co-occurring failure type.

\subsection{From predictive models to pricing}
Based on the calibrated predictive models, the break-even price $P^\star$ in Equations \eqref{eq: breakeven P} and \eqref{eq: cond break even price}, can now be determined.
We make this explicit by reformulating the break-even price in function of the predictive models, taking into account the risk factors:
\begin{equation}
\begin{aligned}
P^\star&\vert \bm{\chi_1},\bm{\chi_2}(t),\bm{w},\bm{w}_y,\bm{w}_c,\bm{z}=E[N_{f,c}\vert\bm{\chi}_1,\bm{\chi}_2(t)]\cdot E[X_{c}\vert \boldsymbol{w}_c] \\ &+E[N_{f,\text{minor}}\vert \bm{\chi}_1,\bm{\chi}_2(t)] \cdot \sum_{y = 1}^{n_{f,\text{minor}}}\Pr[Y=y \vert Y\in \mathcal{N}_{f,\text{minor}} ,\bm{z}]\cdot E[X_{y}\vert \boldsymbol{w}_y]\\ &+  n_{m} \cdot E[S\vert \boldsymbol{w}],
\end{aligned}	
\label{eq: price (pricing)}
\end{equation}
where $E[N_{f,\text{minor}}\vert\bm{\chi}_1,\bm{\chi}_2(t)]$ is the predicted number of minor failures within the contract duration and $E[N_{f,c}\vert \bm{\chi}_1,\bm{\chi}_2(t)]$ the predicted number of catastrophic failures for machine characteristics $\bm{\chi}_1$ and $\bm{\chi}_2(t)$, obtained from the predictive time-to-failure models with likelihood given in Equation~\eqref{eq: lik}.
Remark that the explicit dependence of the price on the contract duration is omitted from Equation \eqref{eq: price (pricing)}, however it is evident that its components are dependent on the contract duration.
As the break-even price depends on the machine characteristics, different tariff plans are generated depending on the set of characteristics that are included in the predictive models. 
As the contract portfolio of the service provider grows, or changes, the models should be re-calibrated to learn from this new data.

Note that the estimates from the predictive models could also be used for other purposes than pricing.
An example of this would be data-driven maintenance optimization, where the predictive models for the failure intensity could serve as an input for the optimization of the preventive maintenance interval.

\iffalse
We opt for Monte Carlo simulation to obtain the estimate the different terms in Equation~\eqref{eq: price (pricing)}.
The generated paths rely on the simulation engine described in Section~\ref{simulation}, where the input parameters for the simulation are the parameters resulting from the calibration of the predictive models.
Next to an estimate for the expected values in Equation~\eqref{eq: price (pricing)}, other risk measures could be estimated from the generated scenarios. 
This allows to estimate a risk margin on top of the break-even price, as well as more detailed risk assessment of the contracts in the portfolio.
\fi

\FloatBarrier
\section{Simulation engine and numerical experiment}
\label{simulation}
\subsection{Motivation and properties}
\label{simulation: motivation}
We develop an engine to generate maintenance and failure data for a portfolio of machines reflecting their heterogeneity induced by the different risk factors.
This engine allows testing the performance and adequacy of our analytic models \citep{bender2005generating, metcalfe2006importance, montez2017simulating}.
Simulated data provide a controlled environment to study the performance of the methodology in different scenarios preparatory to applying it to real data.
At the same time, simulated data provide a recipe for data collection since real data (after cleaning) should have a similar format and features as the simulated data.

The dataset used for pricing should contain the time stamps of failures, including their type and  associated costs, the time of preventive maintenance and their costs, together with the machine characteristics.
Simulating event occurrences and their associated impact (as in Figure~\ref{fig: machine life}) is not straightforward, as extensively documented in the biostatistical literature where the recurrence of events during an individual's lifetime is studied \citep{metcalfe2006importance,beyersmann2009simulating,hendry2014data,jahn2015simulating,penichoux2015simulating}.
The simulation engine should be capable of handling the following list of analytic ingredients.
First, the engine simulates the occurrence of recurrent events over time \citep{cook2007statistical,penichoux2015simulating} to generate multiple occurrences of the same type of failure.
Next, the engine generates occurrences of different types of events \citep{beyersmann2009simulating,frees2008hierarchical} of which some can be terminal, i.e. equivalent to death in the biostatistical setting.
%\citep{beyersmann2009simulating} \citep{frees2008hierarchical} 
The data should also include machine characteristics, which can be time-independent \citep{metcalfe2006importance}, e.g. the country of residence, operational environment, operator and maintenance crew skill level, or time-dependent \citep{hendry2014data}, e.g. temperature, maintenance and service history, which impact failure times, types and costs, and maintenance costs.
To introduce non-observable heterogeneity, i.e.\ not reflected by the observed machine characteristics, the engine could be extended with the inclusion of frailties or random effects \citep{penichoux2015simulating}.
Machines may have exactly the same risk factors, but at the same time some of them might fail significantly more.
The inclusion of frailties creates additional complexity; since our focus is on observable risk factors, we have not included them in our model.
Finally, risk-free intervals \citep{jahn2015simulating}, during which no failures can occur, may account for the actual duration of a maintenance intervention.

\subsection{Set up of the engine}
\label{sec: simulation model}
We generate data for each machine $i\ (\in \{1,2,...,n\})$ in a portfolio of $n = 5\ 000$ machines, during an observation time $t_{\text{obs},i}$.
These machines could be of different types, in which case the machine type should be added as an extra covariate.
In our numerical experiment we will assume that all machines have the same type.
The observation time of each machine is either 5 years or uniformly distributed between three and five years.
We simulate each machine in a time window $[t_{0,i},t_{0,i} + t_{\text{obs},i}]$. 
Without loss of generality we will consider $t_{0,i} = 0$ for all machines.
%\st{Remark that $t_{\text{obs},i}$ can also be longer than $t_{\text{contract}}$ in case the machine is observed during a period it was not under a contract, however this case is not implemented in our simulation. All failures that would occur outside the observation window are censored.}
Each machine~$i$ has a number of characteristics or covariates $\bm{x}_i\ (= \bm{x}_{1,i} \cup \bm{x}_{2,i}(t))$, where we make a distinction between time-independent $\bm{x}_{1,i}$ and time-dependent $\bm{x}_{2,i}(t)$ covariates.
%The covariates $\bm{x}$ should be considered as the full set of covariates which are to our disposal.
%Subsets of this set will influence the different components of the simulation engine. \colr{Laurens: wat bedoel je met deze laatste zin?}\todo{dat niet alle covariaten alle delen van de simulatie beinvloeden}.
Each machine has 4 fixed covariates, $\bm{x}_1 = (x_{1,1},x_{1,2},x_{1,3},x_{1,4})^T \in \{0,1\}^4$.
For the purpose of simulation, we randomly allocate the values with equal probability to the fixed covariates.
We adhere to a general specification of the fixed covariates, while in reality these will collect
machine-specific risk factors as outlined in Section \ref{sec: introduction}.
We let $x_2(t)$ indicate the time-varying covariate whether a type 3 minor failure has occurred,
\begin{equation}
x_2(t) = 
\begin{cases}
	 &0 \text{, if no type 3 minor failure has occurred before $t$} \\
	&1 \text{, if at least one type 3 minor failure has occurred before $t$}.
\end{cases}
\end{equation}
The time-varying covariate $x_2(t)$ can be seen as a collateral damage of a type 3 failure: i.e.
if a type 3 failure occurs, the overall reliability of the machine deteriorates. Other time-varying covariates could represent e.g.\ the time since last maintenance event, the number of maintenance events in the past year, the cost of the last preventive maintenance, etc.

Next, we determine the number of failure types $n_f = 1 + n_{f,\text{minor}}$, with $n_{f,\text{minor}}$ the number of minor failure types, and we additionally consider one catastrophic failure.
We consider $n_{f,\text{minor}} = 3$ minor failure types (type 1, type 2 and type 3), of which type 3 is the co-occurrence of failures of type 1 and 2.
The number of periodic maintenance actions during the contract duration is determined by fixing the maintenance interval $\Delta t_{\text{PM}} = 1$.
In our experiment, we let the covariates only impact the distribution of the time-to-failure and failure types, but not the costs, although that should not necessarily be the case.
Consequently $\bm{\chi}_1 = \bm{x}_1$ and $\bm{\chi}_2(t) = x_2(t)$, $\bm{z} = \bm{x}_1$, and $\bm{w} = \emptyset, \bm{w}_y = \emptyset$ and $\bm{w}_c = \emptyset$ in Equations \eqref{eq: minor failure hazard}, \eqref{eq: multinom prob} and \eqref{eq: costs distr} respectively.
An overview of the simulation parameters can be found in Table \ref{tab: pricing} in Appendix \ref{appendix: param en est}.

\vspace{5mm}
\noindent
We provide the pseudo-code of our simulation engine in Algorithm \ref{alg: simulation}, see Appendix~\ref{appendix: algo}.
In what follows we describe how the algorithm generates for each machine~$i$ failures (times, types and costs), and periodic maintenance actions (and their associated costs) within the observation time $t_{\text{obs},i}$.
%Finally, the time-line of machine $i$ is included in the portfolio dataset.
\paragraph{Failure times}
Failure times are generated independent of the failure type by the function {\tt getFailureTime($\bm{x}_i$,$t_{\text{previous}}$)}, where $\bm{x}_i\ (= \bm{x}_{1,i} \cup \bm{x}_{2,i}(t))$ is the covariate vector of machine~$i$ and $t_{\text{previous}}$ the event time of the previous failure on machine~$i$. 
We simulate the failure times making use of failure intensities, see Equations \eqref{eq: minor failure hazard} and \eqref{eq: weibull hazard} for minor and catastrophic failures respectively. 
Recurrent minor failure times as well as a catastrophic (terminal) failure are generated from $\lambda_{\text{total}}(t)$ making use of inverse transform sampling  \citep{metcalfe2006importance,cook2007statistical,jahn2015simulating,penichoux2015simulating}.
To facilitate inverse transform sampling, the time-varying covariates ${x}_{2,i}(t)$ should be piecewise constant functions.
The latter is the case for our simulation.
\paragraph{Failure type}
For each failure with failure time $t$,  the function {\tt getFailureType($\bm{x}_i$,$t$)} generates the failure type $Y$ through a two-step process, where $\bm{x}_i$ is the covariate vector of machine~$i$.
First, we determine if the failure is minor or catastrophic.
These failure types occur with probabilities \citep{beyersmann2009simulating},
\begin{equation}
\begin{aligned}
\Pr[Y = f_m] &= \frac{\lambda_{f_m}(t)}{\lambda_{\text{total}}(t)}\\
\Pr[Y = f_c] &= \frac{\lambda_{f_c}(t)}{\lambda_{\text{total}}(t)},
\end{aligned}
\end{equation}
for a failure at time $t$.
Minor and catastrophic failures cannot occur at the same time in the current specification of the model.
However, the simulation engine could be extended by introducing an extra event-type capturing the co-occurrence of a minor failure and catastrophic failure with its own specific failure intensity.

Second, the type of a minor failure, i.e. $\{f_1,f_2,...,f_{n_{f,\text{minor}}}\}$, is generated by sampling from a multinomial distribution \citep{frees2008hierarchical} with the probabilities of Equation \eqref{eq: multinom prob}.
Multiple coinciding minor failures are modelled as a single minor failure type, in our set-up type 3 is the co-occurrence of a type 1 and a type 2 failure.

\paragraph{Costs of failure and maintenance}
We generate $S_j$, the costs of the $j$th maintenance, and $X_{y,k}$, the costs of the $k$th failure of type $y$, with the functions {\tt getMaintenanceCosts($\bm{x}_i$)} and {\tt getFailureCosts($\bm{x}_i,y$)}.
Hereto, we simulate from a positive and right-skewed distribution, i.e. a gamma distribution.
The specifications of the distributions were introduced in Equation \eqref{eq: costs distr}.
A copula with positive dependence, i.e. the Frank copula, captures the statistical dependence between the costs of minor failures occurring at the same time \citep{frees2008hierarchical}.

\begin{figure}[h]
	\centering
	\begin{tikzpicture}
	\definecolor{lightb}{RGB}{153,204,255}
	\definecolor{dm}{RGB}{222, 89, 113}
	\definecolor{m}{RGB}{255,0,255}
	\draw [fill = black, black] (0,-0.05) rectangle (0.1,0.05) {};
	\node [right] at (0.1,0.04) {\small{maintenance}};
	\draw [fill = lightb, lightb] (2.4,-0.05) rectangle (2.5,0.05) {};
	\node [right] at (2.5,0.01) {\small{catastrophic failure}};
	\end{tikzpicture}
	
	\begin{tikzpicture}
	\definecolor{lightb}{RGB}{153,204,255}
	\definecolor{dm}{RGB}{222, 89, 113}
	\definecolor{m}{RGB}{255,0,255}
	\draw [fill = red, red] (5.6,-0.05) rectangle (5.7,0.05) {};
	\node [right] at (5.7,0.01) {\small{failure type 1}};
	\draw [fill = m, m] (8.0,-0.05) rectangle (8.1,0.05) {};
	\node [right] at (8.1,0.01) {\small{failure type 2}};
	\draw [fill = dm, dm] (10.5,-0.05) rectangle (10.6,0.05) {};
	\node [right] at (10.6,0.01) {\small{failure type 3}};
	\end{tikzpicture}
	\includegraphics[width = \textwidth]{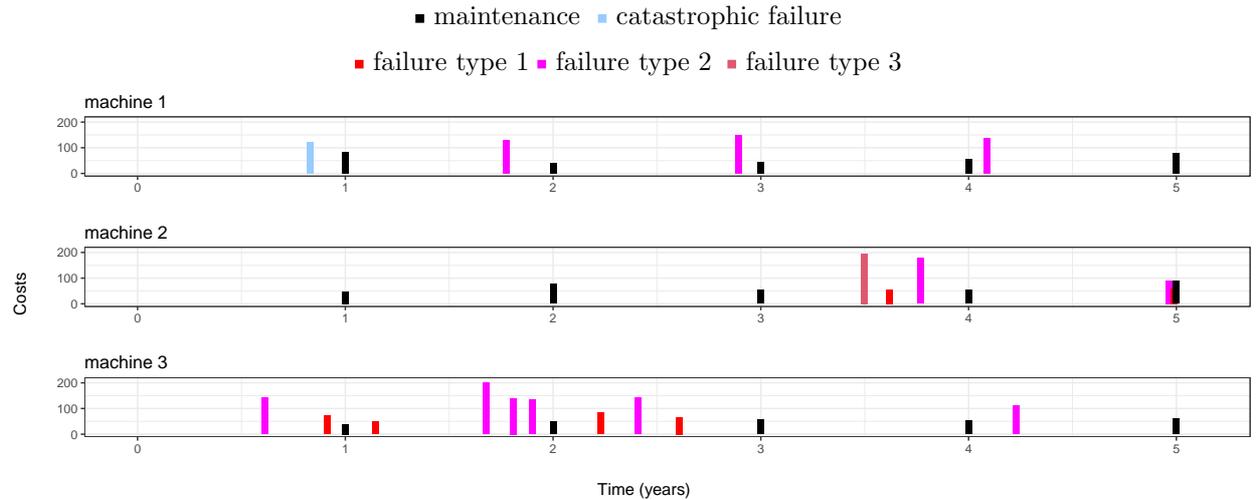}
	\caption{Simulated time-lines for three machines during an observation length of (at most) five years}
	\label{fig: timelines}
\end{figure}

\FloatBarrier
\vspace{5mm}
\noindent
Figure~\ref{fig: timelines} illustrates the generated event time-lines of three arbitrary machines, which result from our simulation engine.

%Note that the calibrated parameters for our generated datasets should be reasonably close to our simulation parameters.
\FloatBarrier

\subsection{Data exploration}
%We acknowledge that this may seem artificial for a simulated dataset, but this step is instrumental when working with real data.

\paragraph{Empirical distribution of exposure and covariates}
We first explore the distribution of the \textquoteleft exposure(-to-risk)\textquoteright\ of our simulated contract portfolio \citep{denuit2007actuarial}.
This quantifies the risk each machine in the dataset is exposed to, expressed in terms of its observation time~$t_{\text{obs}}$. 
Although we use time as the unit of exposure, machine run-time could equally be used.
The relative exposure for each machine $i$ is 
\begin{equation}
    \text{relative exposure}_i = \frac{\text{exposure of machine } i}{\text{total exposure}} = \frac{t_{\text{obs},i}}{\sum_{i = 1}^{n} t_{\text{obs},i}}.
\end{equation}
Summing the relative exposures for machines taking the same values in the risk factors leads to Figure~\ref{fig: total expo}.
Figure~\ref{fig: total expo} (top left panel) illustrates that for our dataset, more than $80\%$ of the total exposure is represented by machines for which five years of data is available.
Figure~\ref{fig: total expo} (right panels) shows that each fixed covariate in $\bm{x}_1$ has equal exposure in each of its values in the dataset, whereas the type 3 minor failure (represented by the time-varying covariate $x_2(t)$) is only observed (at least once) in less than $20\%$ of the total exposure. 
\iffalse
\begin{figure}[tb]
	\centering
	\begin{minipage}{0.24\textwidth}
		\centering
		\includegraphics[width = \textwidth]{figures/expo_1}
		\caption*{\small{contract}}
	\end{minipage}
	\centering
	\hfill
	\begin{minipage}{0.24\textwidth}
		\centering
		\includegraphics[width = \textwidth]{figures/expo}
		\caption*{\small{fixed covariates}}
	\end{minipage}
	\hfill
	\centering
	\begin{minipage}{0.24\textwidth}
		\centering
		\includegraphics[width = \textwidth]{figures/expo_tvc}
		\caption*{\small{time-varying covariate}}
	\end{minipage}
\caption{Relative exposure of the machine portfolio: the majority of the machines have data for five years (left), the fixed covariates~$\bm{x_1}$ are equally represented (middle) and the time-varying covariate~$x_2(t)$ has value 1 in less than $20\%$ of the machine portfolio (right).}
\label{fig: total expo}
\end{figure}
\fi
\begin{figure}[tb]
	\centering
	\includegraphics[width = \textwidth]{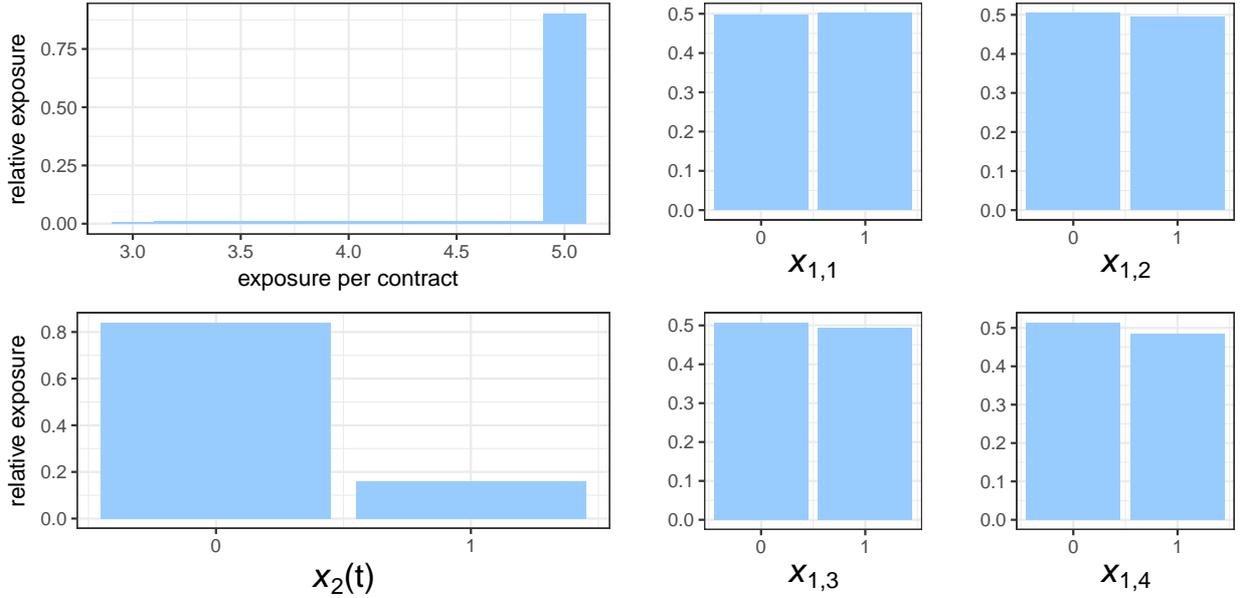}
    \caption{Relative exposure of the machine portfolio per contract (top left), in the time-varying covariate~$x_2(t)$ (bottom left) and in the fixed covariates~$\bm{x_1}$ (right)}
\label{fig: total expo}
\end{figure}

\begin{figure}[tb]
\centering
\includegraphics[width = \textwidth]{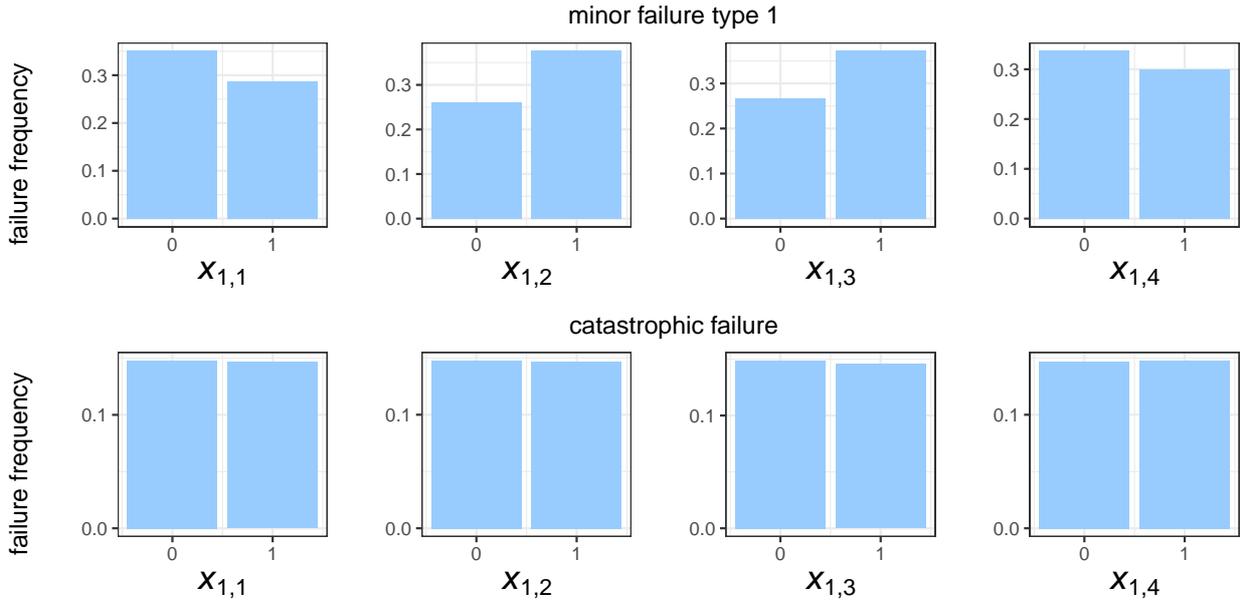}
\caption{Empirical failure frequency for type 1 minor and catastrophic failures as a function of the observed covariates.}
\label{fig: emp intensities}
\end{figure}

\paragraph{Influence of the covariates}
We explore the influence of the covariates on the number of failures by picturing the empirical failure frequency for specific machine profiles, e.g. machines with $x_{1,1} = 1$. 
This empirical failure frequency is expressed as the ratio of the number of failures of a specified type and the total exposure, both calculated for the machine profile under consideration.
Taking the ratio is necessary to take the exposure (observation time) into account, since more failures are expected when a machine profile is observed for a longer period.
Figure \ref{fig: emp intensities} illustrates how the covariates influence the failure behaviour for the minor failures, illustrated for minor failure type 1, but have no influence on the catastrophic failures.

\FloatBarrier
\subsection{Calibration of the predictive models and resulting pricing schemes}
\label{sec: verify the sim}
Table~\ref{tab: pricing} (see Appendix \ref{appendix: param en est}) provides the estimated parameters and their $95\%-$confidence intervals resulting from the calibration of the predictive models using the calibration strategy outlined in Section~\ref{sec: calibration}. 
The confidence intervals follow from the properties of maximum likelihood estimators.
%A verification of our simulation engine by simulating and calibrating multiple datasets is provided in Appendix~\ref{appendix verification}.

The choice (or availability) of the machine-dependent covariates $\bm{\chi}_1,\bm{\chi}_2(t)$, $\boldsymbol{z}$, $\boldsymbol{w},\boldsymbol{w}_{y}$ and $\boldsymbol{w}_{c}$ determines the tariff plan. 
For a real-life dataset, the relevant covariates have to be obtained by means of variable selection strategies \citep{henckaerts2018data}.
For the purpose of our experiment, however, we assume all covariates are relevant and omit the selection step.
In what follows, we introduce three different tariff plans with respective break-even prices $P_a^\star, P_b^\star$ and $P_c^\star$. In tariff plan~$a$ we do not include any covariates, which means there is no price differentiation and all customers pay the same price. Tariff plan~$b$ only takes the fixed covariates into account, and tariff plan~$c$ takes both the fixed and time-dependent covariates into account, see Table~\ref{tab: covariate sets}.
%In Section~\ref{econ value} we will evaluate these pricing schemes.
We calibrate the predictive models for each combination of covariates taken into consideration.
\FloatBarrier
\begin{table}[tb]
\caption{The set of covariates takes into account for the tariff plans $a$, $b$ and $c$.}
\label{tab: covariate sets}
	\centering
\begin{tabular}{lcccc}
	\toprule
	& & $a$ & $b$ & $c$\\
	\midrule
	failure time & $\boldsymbol{\chi}_1$ &$\emptyset$&$\bm{x}_1$& $\bm{x}_1$ \\
	 &$\bm{\chi}_2(t)$ &$\emptyset$&$\emptyset$& $x_2(t)$ \\
	failure types & $\boldsymbol{z}$ &$\emptyset$&$\bm{x}_1$&$\bm{x}_1 $ \\
	costs & $\boldsymbol{w},\boldsymbol{w}_{y},\boldsymbol{w}_{c}$ &$\emptyset$& $\emptyset$ & $\emptyset$\\ 
%	$\boldsymbol{w}_{2}$ &$\emptyset$&$\emptyset$ &$\emptyset$\\
%	$\boldsymbol{w}_{3}$ &$\emptyset$&$\emptyset$ &$\emptyset$\\
	\bottomrule
\end{tabular}
\end{table}
For each machine profile, Figure~\ref{tab: price list} displays the resulting break-even prices prescribed by the different tariff plans for a contract duration of two years.
The different machine profiles arise from the four fixed covariates in vector $\bm{x}_1$, taking two different levels each.
Additionally, a machine either has experienced a minor failure of type $3$ before the start of the contract, $x_2(t=0) = 1$, or not, $x_2(t=0) = 0$. 
This leads to $32$ different machine profiles upon initiation of the contract, resulting in $32$ different prices under tariff plan $c$.
Tariff plan $b$ does not take into account the time-varying covariate $x_2(t)$ and as such it generates only $16$ different prices.
Tariff plan $a$ returns the same price for every machine profile since it does not consider any risk factors.

Although the average, over all machine profiles, break-even price under all tariff plans is very similar, the difference between the price under tariff $a$ and the prices under tariffs $b$ and $c$ can be quite significant.
This effect arises because tariff $a$ overcharges low risk machines and undercharges high risk machines. 
The difference between tariff $b$ and $c$ is less pronounced since the influence of the time-varying covariate is minor.
It is necessary to stress that the price of a one-year contract is not half of the two-year contract prices displayed in Figure~\ref{tab: price list}, nor is the price of a four-year contract double the price of a two-year contract.
The price of the contracts is non-homogeneous in time since the failure intensity is non-homogeneous.

\iffalse
\begin{table}[t]
	\caption{Price list (contract duration = 5 years) for the different pricing schemes for each machine profile}
	\label{tab: price list}
	\centering
	\begin{tabular}{cccc}
		\toprule
		$\bm{x}_1$ & $P^\star_a$ & $P^\star_b$ & $P^\star_c$\\
		\midrule
		$(0, 0, 0, 0)$ & $930.34$ & $826.72$ & $825.98$\\
		$(1, 0, 0, 0)$ & $930.34$ & $752.67$ & $749.85$\\
		$(0, 1, 0, 0)$ & $930.34$ & $958.64$ & $957.12$\\ 
		$(0, 0, 1, 0)$ & $930.34$ & $1028.03$ & $1028.70$\\
		$(0, 0, 0, 1)$ & $930.34$ & $790.25$ & $788.12$\\
		$(1, 1, 0, 0)$ & $930.34$ & $860.39$ & $857.24$\\
		$(1, 0, 1, 0)$ & $930.34$ & $917.14$ & $918.04$\\
		$(1, 0, 0, 1)$ & $930.34$ & $721.92$ & $721.08$\\ 
		$(0, 1, 1, 0)$ & $930.34$ & $1233.78$ & $1234.89$\\
		$(0, 1, 0, 1)$ & $930.34$ & $905.23$ & $905.70$\\
		$(0, 0, 1, 1)$ & $930.34$ & $972.26$ & $970.26$\\
		$(1, 1, 1, 0)$ & $930.34$ & $1082.29$ & $1081.68$\\
		$(0, 1, 1, 1)$ & $930.34$ & $1156.04$ & $1153.77$\\
		$(1, 0, 1, 1)$ & $930.34$ & $873.74$ & $872.32$\\ 
		$(1, 1, 0, 1)$ & $930.34$ & $818.38$ & $817.53$\\ 
		$(1, 1, 1, 1)$ & $930.34$ & $1024.00$ & $1021.45$\\
		average		& $930.34$ & $932.59$ & $931.48$\\
		\bottomrule
	\end{tabular}
\end{table}
\fi
\begin{figure}[h]
\centering
	\begin{tikzpicture}
	\definecolor{lightb}{RGB}{153,204,255}
	\definecolor{dm}{RGB}{222, 89, 113}
	\definecolor{m}{RGB}{255, 0, 255}
	\draw [fill = lightb, lightb] (5.6,-0.05) rectangle (5.7,0.05) {};
	\node [right] at (5.7,0.01) {\small{tariff a}};
	\draw [fill = m, m] (8.0,-0.05) rectangle (8.1,0.05) {};
	\node [right] at (8.1,0.01) {\small{tariff b}};
	\draw [fill = red, red] (10.5,-0.05) rectangle (10.6,0.05) {};
	\node [right] at (10.6,0.01) {\small{tariff c}};
\end{tikzpicture}
\centering
\includegraphics[width = \textwidth]{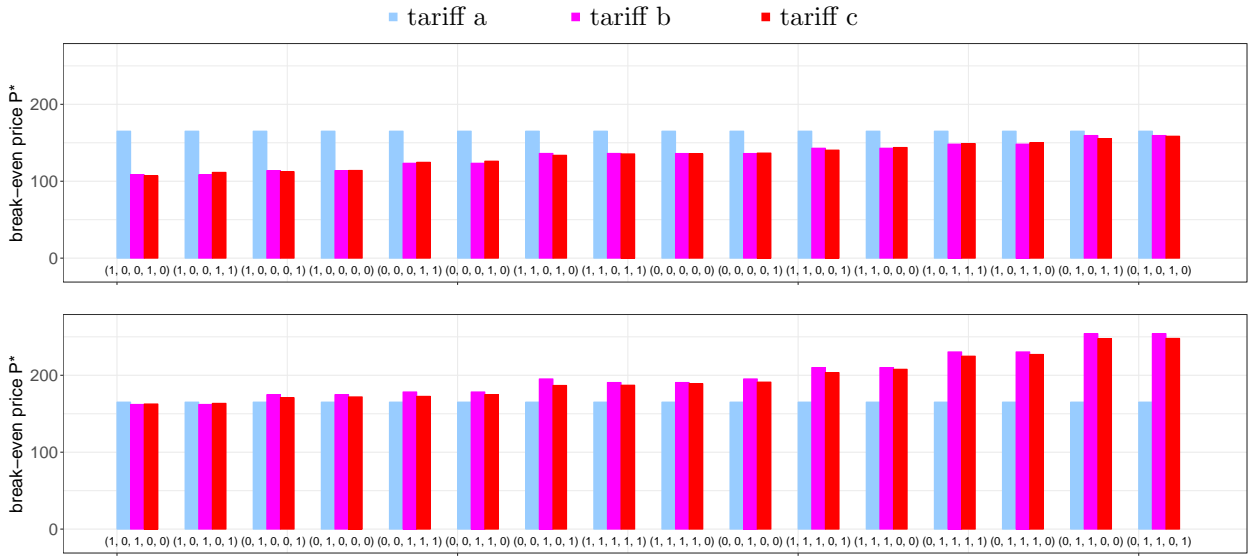}
\caption{Break-even prices $P^\star$ under the different tariff plans for each machine profile $(x_{1,1},x_{1,2},x_{1,3},x_{1,4},x_{2}(t=0))$ for a two-year contract.}
\label{tab: price list}
\end{figure}

In Appendix~\ref{appendix verification} (Supplementary material) we provide the results of an extensive simulation study, where we generate multiple portfolios of machines over time (along the recipe in Section \ref{sec: simulation model}) and calibrate the proposed hierarchical likelihood on each of these simulated datasets. 
The purpose of this simulation study is to verify the accurracy of the calibration strategy described in Section \ref{sec: calibration}.
\FloatBarrier
\section{Model lift: Evaluation of different tariff plans}
\label{econ value}
\begin{figure}[tb]
	\centering
	\begin{tikzpicture}[scale=0.8,snake=zigzag, line before snake = 5mm, line after snake = 5mm]
	% axis
	\draw [-] (0,0)--(7.5,0);
	\draw [dashed] (7.6,0)--(13,0);
	\node at (0,0) {\tiny{$\bullet$}};
	\node [below] at (0,-.1) {$0$};
	
	\draw [-] (3,-2)--(7.5,-2);
	\draw [dashed] (7.6,-2)--(13,-2);
	\node at (3,-2) {\tiny{$\bullet$}};
	\node [below] at (3,-2.1) {$0$};
	
	\draw [-] (0,-4.5)--(7.5,-4.5);
	\draw [dashed] (7.6,-4.5)--(13,-4.5);
	\node at (0,-4.5) {\tiny{$\bullet$}};
	\node [below] at (0,-4.6) {$0$};
	
	\draw [-] (0,-6)--(7.5,-6);
	\draw [dashed,->] (7.6,-6)--(14,-6);
	\node [right] at (14,-6) {time};
	\node at (7.5,-6) {\tiny{$\bullet$}};
	\node [below] at (7.5,-6.1) {$t^\star$};
	% titles
	\node [left] at (-0.5,0) {machine 1};
	\node [left] at (-0.5,-2) {machine 2};
	\node at (-1,-3) {\tiny{$\bullet$}};
	\node at (-1,-3.25) {\tiny{$\bullet$}};
	\node at (-1,-3.5) {\tiny{$\bullet$}};
	\node [left] at (-0.5,-4.5) {machine $n$};
	\node [above] at (3.5,0.5) {Training: in-time};
	\node [above] at (10.5,0.5) {Test: out-of-time};
	
	%	\visible<1>{\draw (5,-7.1) node {Split dataset in 2 parts: {\color{black}in-time} and {\color{red}out-of-time} $\ldots$};}
	%	\visible<2>{\draw (5,-7.1) node {$\ldots$ use the {\color{black}in-time} data for training $\ldots$};}
	%	\visible<3->{\draw (5,-7.1) node {$\ldots$ use the {\color{red}out-of-time} data for testing.};}
	
	% boxes
	\draw [rounded corners,thick,opacity = 0.5] (-.5,0.5) rectangle (7.5, -5.5) {};
	\draw [rounded corners,thick,opacity = 0.5,dashed] (7.5,0.5) rectangle (13.5, -5.5) {};
	
	% maintenance
	\node [black,below,font=\sffamily] at (4,-2.1) {$m$};
	\node [black,below] at (6,-2.1) {$m$};
	\node [black,below] at (8,-2.1) {$m$};
	\node [black,below] at (10,-2.1) {$m$};
	\node [black,below] at (12,-2.1) {$m$};
	\node [black] at (4,-2) {\tiny{$\bullet$}};
	\node [black] at (6,-2) {\tiny{$\bullet$}};
	\node [black] at (8,-2) {\tiny{$\bullet$}};
	\node [black] at (10,-2) {\tiny{$\bullet$}};
	\node [black] at (12,-2) {\tiny{$\bullet$}};
	
	\node [black,below,font=\sffamily] at (2,-.1) {$m$};
	\node [black,below] at (4,-.1) {$m$};
	\node [black,below] at (6,-.1) {$m$};
	\node [black,below] at (8,-.1) {$m$};
	\node [black,below] at (10,-.1) {$m$};
	\node [black,below] at (12,-.1) {$m$};
	\node [black] at (2,0) {\tiny{$\bullet$}};
	\node [black] at (4,0) {\tiny{$\bullet$}};
	\node [black] at (6,0) {\tiny{$\bullet$}};
	\node [black] at (8,0) {\tiny{$\bullet$}};
	\node [black] at (10,0) {\tiny{$\bullet$}};
	\node [black] at (12,0) {\tiny{$\bullet$}};
	
	\node [black,below,font=\sffamily] at (2,-4.6) {$m$};
	\node [black,below] at (4,-4.6) {$m$};
	\node [black,below] at (6,-4.6) {$m$};
	\node [black,below] at (8,-4.6) {$m$};
	\node [black,below] at (10,-4.6) {$m$};
	\node [black,below] at (12,-4.6) {$m$};
	\node [black] at (2,-4.5) {\tiny{$\bullet$}};
	\node [black] at (4,-4.5) {\tiny{$\bullet$}};
	\node [black] at (6,-4.5) {\tiny{$\bullet$}};
	\node [black] at (8,-4.5) {\tiny{$\bullet$}};
	\node [black] at (10,-4.5) {\tiny{$\bullet$}};
	\node [black] at (12,-4.5) {\tiny{$\bullet$}};
	
	% failure type 1
	\node [red,below,font=\sffamily] at (3,-.1) {$f_1$};
	\node [red,below] at (9,-.1) {$f_1$};
	\node [red] at (3,0) {\tiny{$\bullet$}};
	\node [red] at (9,0) {\tiny{$\bullet$}};
	
	\node [red,below,font=\sffamily] at (12.5,-2.1) {$f_1$};
	\node [red] at (12.5,-2) {\tiny{$\bullet$}};
	\node [red,below] at (5,-2.1) {$f_1$};
	\node [red] at (5,-2) {\tiny{$\bullet$}};
	
	\node [red,below,font=\sffamily] at (3.1,-4.6) {$f_1$};
	\node [red,below] at (9,-4.6) {$f_1$};
	\node [red] at (3.1,-4.5) {\tiny{$\bullet$}};
	\node [red] at (9,-4.5) {\tiny{$\bullet$}};
	\node [red,below] at (5.2,-4.6) {$f_1$};
	\node [red] at (5.2,-4.5) {\tiny{$\bullet$}};
	
	% failure 2
	\node [magenta,below] at (7,-.1) {$f_2$};
	\node [magenta] at (7,0) {\tiny{$\bullet$}};
	
	\node [magenta,below] at (8.6,-2.1) {$f_2$};
	\node [magenta] at (8.6,-2) {\tiny{$\bullet$}};
	
	\node [magenta,below] at (1.6,-4.6) {$f_2$};
	\node [magenta] at (1.6,-4.5) {\tiny{$\bullet$}};
	\node [magenta,below] at (4.4,-4.6) {$f_2$};
	\node [magenta] at (4.4,-4.5) {\tiny{$\bullet$}};
	\node [magenta,below] at (11.6,-4.6) {$f_2$};
	\node [magenta] at (11.6,-4.5) {\tiny{$\bullet$}};
	\end{tikzpicture}
	\caption{The data is split in an in-time training set to calibrate the predictive models, and an out-of-time set to evaluate the calibrated models.}
	\label{fig: in and out of sample}
\end{figure}
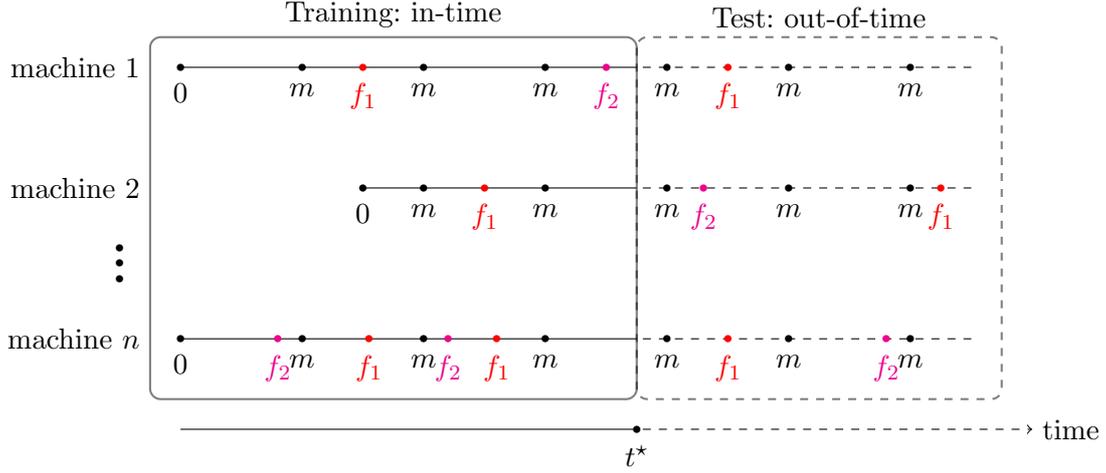

We evaluate the different tariff plans $P_a^\star, P_b^\star$ and $P_c^\star$ introduced in Table~\ref{tab: covariate sets} for our simulated portfolio of $5000$ machines by comparing the collected service fees, with the actual, out-of-time collected costs.
We divide the machine data in two parts: an in-time training set to calibrate the predictive models and to determine the prices, and an out-of-time set, which returns the actual costs \citep{goldburd2016generalized}, see Figure~\ref{fig: in and out of sample}.
With our simulation engine we generate both in-time data (with an observation time of maximum five years), as well as out-of-time data (five extra years data for each machine) with the same simulation settings.
We not only measure the accuracy of the tariff plan, i.e. the extent to which the tariff plan (calibrated in-time) covers the actual costs (as observed out-of-time) in the portfolio, but also its capability to identify machine heterogeneity and adapt prices accordingly.

\subsection{Accuracy of tariff plans}
\label{sec: accuracy}
\begin{figure}[tb]
\centering
\includegraphics[width = \textwidth]{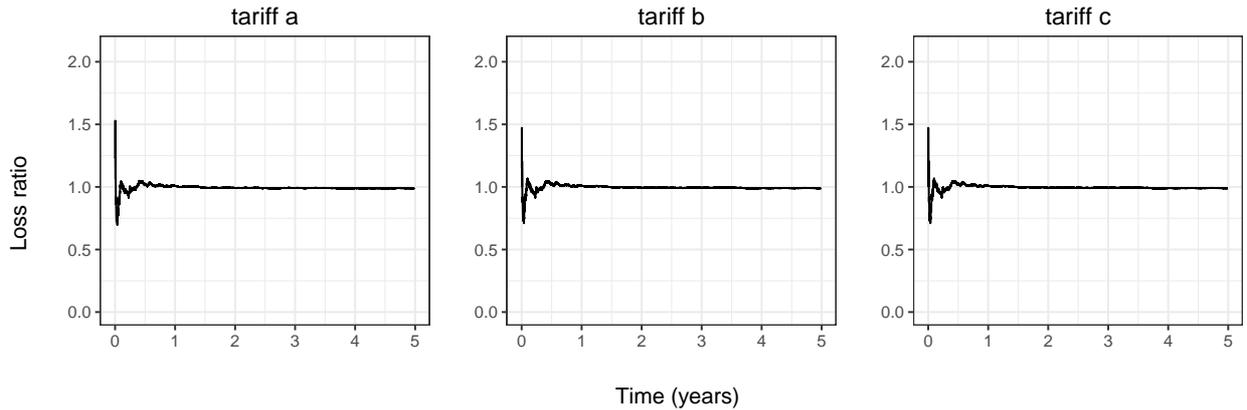}
\caption{The evolution of the out-of-time loss ratio over time.}
\label{fig loss ratio evolution}
\end{figure}

The loss ratio for a given tariff plan is a measure for tariff accuracy. 
It is defined as the ratio of the total actual costs realized in the entire contract portfolio over the total collected services fees under the tariff plan.
As such, we can express the loss ratio as follows,
\begin{equation}
    \text{loss ratio} = \frac{\sum_{i=1}^{n} C_i(\Delta t)}{\sum_{i=1}^{n} P^\star_i(\Delta t)},
\end{equation}
where the sum is over all contracts $i$ in the portfolio.
Table~\ref{tab: loss ratios} confirms that over a horizon of two years, the three considered tariff plans all have an out-of-time loss ratio close to one. 
Figure~\ref{fig loss ratio evolution} plots the out-of-time loss ratio over time, confirming Table~\ref{tab: loss ratios} that the loss ratio rapidly stabilizes to a value close to one.
That means that at the (aggregated) level of the contract portfolio, all three tariff plans are break-even. 
Nevertheless, that does not imply that each contract (at the level of the individual machine) is expected to be break-even under the three tariff plans. 

\begin{table}[tb]
\caption{The out-of-time loss ratios for the three tariff plans over a two year horizon.}
\label{tab: loss ratios}
\centering
\begin{tabular}{cccc}
	\toprule
	tariff plan&$a$&$b$&$c$\\
	\midrule
	loss ratio& $0.992$ & $0.989$ & $0.994$\\
	\bottomrule
\end{tabular}
\end{table}

%\st{As such all tariff plans can be considered accurate. Since the three tariff plans have similar behaviour in the loss ratio, a preference for a specific tariff plan can not be based on the loss ratio.}

\FloatBarrier
\subsection{Capability to differentiate}
To measure the capability of a tariff plan to differentiate its prices based on the machine profiles (and their corresponding risk), we use the so-called model lift \citep{goldburd2016generalized}.  
A better capability to differentiate prices implies that the contracts of each machine profile are expected to be break-even, and it prevents adverse selection, as low-risk customers will be offered a lower price than high-risk customers and thus have less incentive to leave. 

\begin{figure}[tb]
    \centering
    \begin{tikzpicture}
	\definecolor{lightb}{RGB}{153,204,255}
	\draw [fill = lightb, lightb] (4.6,-0.05) rectangle (4.7,0.05) {};
	\node [right] at (4.7,0.01) {\small{average costs per bin}};
	\draw [fill = red, red] (8.6,-0.05) rectangle (8.7,0.05) {};
	\node [right] at (8.7,0.01) {\small{average price per bin}};
\end{tikzpicture}
	\centering
	\includegraphics[width = \textwidth]{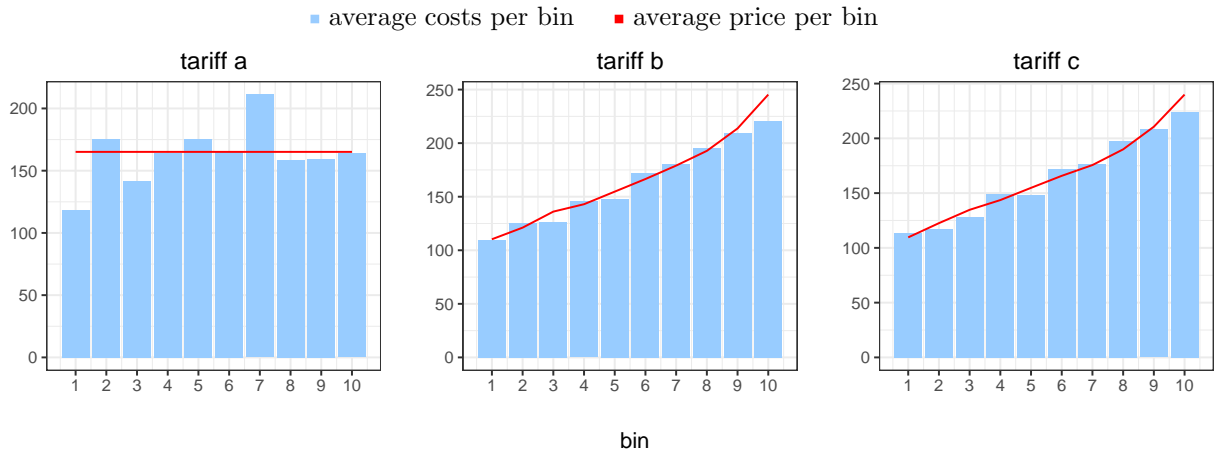}
	\caption{Quantile plot over a two year horizon for the different tariff plans. The height of the bars indicate the average costs per bin and the red line the average price.}
	\label{fig quantile plot}
\end{figure}

\paragraph{Quantile plot}
%Since the price lists does not provide insight in the correctness of the differentiating capacity of the tariff plan.
The quantile plots in Figure~\ref{fig quantile plot} graphically compare the actual, out-of-time, total costs incurred during the contract with the proposed machine-specific break-even price over a two year horizon. 
For a portfolio of $n$ contracts with price $P^\star_i$ for machine $i$, incurred costs $C_i$ and machine characteristics $\bm{x}_i$ $(=\bm{x}_{1,i}\cup \bm{x}_{2,i}(t))$, the quantile plot is obtained by first sorting the contracts in ascending order of price $P^\star_i$ and binning them in equal groups, so-called quantiles; for each bin we then compare the average costs against the corresponding average price.
The quantile plots in Figure~\ref{fig quantile plot} indicate that tariff plans $b$ and $c$ succeed in charging a higher price to the more maintenance-heavy contracts.
The gap between the average price and average costs for each bin is small, indicating that the contracts in each bin are (close to) break-even on average.
Tariff plan $a$ sorts the contracts randomly since all contracts have the same price and consequently each bin has a similar average cost.
This contrasts to tariffs $b$ and $c$, who impose an ordering of the contracts from maintenance-light to maintenance-heavy and as such there is an upward trend in average costs for each quantile.
Clearly, under tariff plan $a$, the average maintenance-light contract is profitable, but the average maintenance-heavy contract is loss-making.

\paragraph{Adverse selection}
The heterogeneity in maintenance costs of the different contracts is identified by tariff plans $b$ and $c$ that take into account the risk factors.
These tariff plans effectively classify contracts based on their risk and distinguish between maintenance intensive (high risk) versus maintenance light (low risk) contracts.
In Figure \ref{fig quantile plot}, we can observe that the first 5 quantiles of contracts for tariff plans $b$ and $c$ have average costs and prices below $150$.
We will call these contracts `low-risk' contracts and we will call the 5 last quantiles `high-risk' contracts.
Since tariff plan $a$ charges all contracts more than $150$, the low-risk customers have an incentive to renege from their contract under tariff $a$.
This would leave the service provider, in case he offers tariff plan $a$ to all customers, with a portfolio of high-risk contracts only.
From Figure~\ref{fig quantile plot}, we can see that these high-risk contracts have higher average costs than the premium charged by tariff plan $a$.
This puts the service provider at risk that he is left with a portfolio that is no longer break-even.
In contrast, tariff plans $b$ and $c$ charge a premium in line with the expected costs of a contract.
Therefore, the customers have less incentive to renege on their contract.
Moreover, when customers leave, this is not detrimental to the profitability since the lost premiums are in line with the costs, which in that case no longer have to be covered by the service provider.

\begin{figure}[tb]
    \centering
	\includegraphics[width = \textwidth]{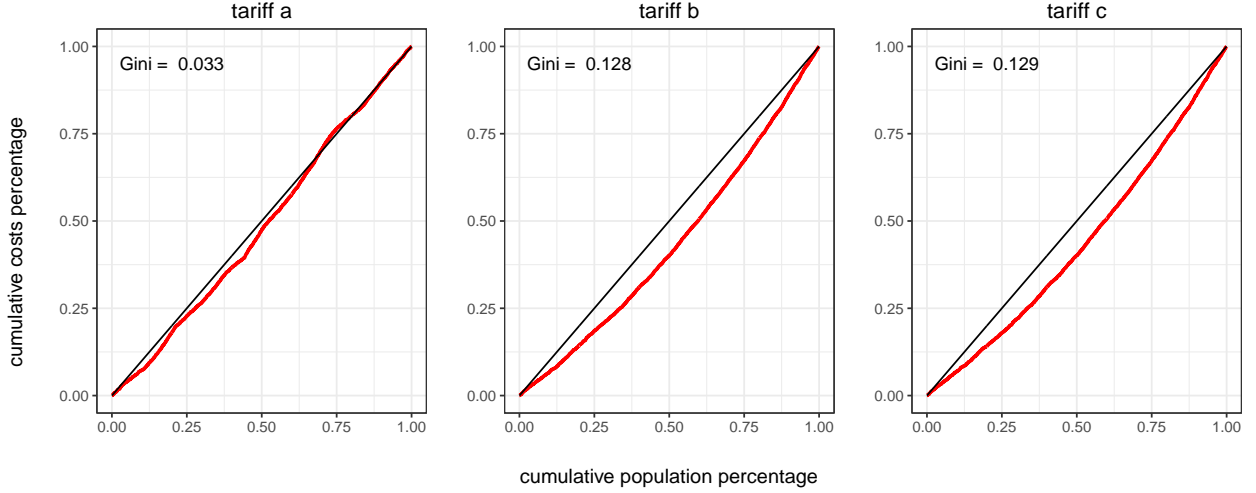}
	\caption{Lorenz curves (red) and Line of Equality (black) for the different tariff plans and costs over two year horizon.}
	\label{fig: Lorenz}
\end{figure}

\paragraph{Lorenz curves and Gini index}
A Lorenz curve \citep{lorenz1905methods} is a tool to compare distributions, frequently used in welfare economics.
We use it here to gain insight in the distribution of incurred costs and prices prescribed by the different tariff plans.
The Lorenz curve is obtained by sorting the machines $i$ in ascending order of price $P^\star_i$, with the cumulative percentage of the machine population on the abscissa and the cumulative percentage of costs $C_i$ on the ordinate \citep{frees2011summarizing,frees2014insurance}. 
The larger the gap between the Lorenz curve and the Line of Equality, i.e., the $45$ degree line, the better the tariff plan's capability
to reflect heterogeneity between machines via a differentiated price list.
The Gini index \citep{gini1912italian} is twice the area between the Lorenz Curve and the Line of Equality.
Figure~\ref{fig: Lorenz} shows the Lorenz curves and corresponding Gini coefficients for the different tariff plans. 
The larger Gini index of tariff plans $b$ and $c$ indicate that they outperform tariff plan $a$ in differentiating between low-risk and high-risk contracts (which is expected since tariff plan $a$ charges every customer the same price).
Remark that we can construct the Lorenz curve for any tariff plan, since it only requires prices and costs as input. 
As such, we could use the Lorenz curve and Gini index to evaluate our pricing methodology against current pricing practices.

\paragraph{Ordered Lorenz curves and Gini index}
As an extension to the Lorenz curves, \citet{frees2011summarizing,frees2014insurance} introduce ordered Lorenz curves. These consider whether an alternative tariff plan $y$ is preferred over the current plan $x$ because it is better capable to differentiate prices.
%Instead of considering the portfolio as $\{(P^\star_i,C_i,\bm{x}_i)\vert\ i \in \{1,...,n\}\}$, we consider two different prices at the same time. 
%This leads to  a portfolio with following specification $\{(P^\star_{x,i},P^\star_{y,i},C_i,\bm{x}_i)\vert\ i \in \{1,...,n\}\}$.
%Where we consider $P^\star_{x,i}$ the current tariff plan of the service provider and $P^\star_{y,i}$ a new tariff plan or the tariff plan of a competitor.
%We want to investigate if it is worthwhile to update the current strategy $P^\star_{x,i}$ to $P^\star_{y,i}$.
The relativity $R_i$ for contract~$i$ is defined by the ratio of the alternative tariff plan over the current plan,
\begin{equation}
	R_i = \frac{P^\star_{y,i}}{P^\star_{x,i}}.
\end{equation}
When both tariff plans cover the expected costs, the average $R_i$ over all contracts~$i$ will be close to one.
A low relativity $R_i < 1$ indicates that price $P^\star_{y,i}$ is lower than $P^\star_{x,i}$, suggesting that tariff plan $y$ is more competitive than tariff $x$ for contract~$i$, and contract~$i$ is potentially lost to a competitor with tariff plan $y$.
A high value $R_i > 1$ for contract~$i$ indicates that $P^\star_{x,i}$ is underpriced and could indicate potential losses when adopting tariff $x$. 
Hence, the more contracts with a relativity differing from 1, the worse tariff plan $x$ is compared to $y$.
This interpretation of the relativities assumes that prices $P^\star_{y}$ are a more accurate representation of the actual costs than~$P^\star_{x}$.
%This leads to the following portfolio $\{(P^\star_{x,i},P^\star_{y,i},C_i,\bm{x}_i, R_i)\vert\ i \in \{1,...,n\}\}$.

To set up the ordered Lorenz curve, the contracts in the portfolio are ordered in increasing values of~$R_i$.
The distribution $F_{P^\star_{x}}(\cdot)$ of the current prices $P^\star_x$ and distribution of the costs $F_{C}(\cdot)$ can then be calculated as:
\begin{equation}
F_{P^\star_{x}}(s) = \frac{\sum_{i=1}^{n} P^\star_{x,i} \ I(R_i\leq s)}{\sum_{i=1}^{n} P^\star_{x,i}}
\hspace{1cm}
F_{C}(s) = \frac{\sum_{i=1}^{n} C_i \ I(R_i\leq s)}{\sum_{i=1}^{n} C_i},
\end{equation}
with $I(\cdot)$ the indicator function, returning 1 if the argument is true and zero otherwise.
The graph \newline $(F_{P^\star_{x}}(s),F_{C}(s))$ then defines the ordered Lorenz curve \citep{frees2011summarizing,frees2014insurance}.
The associated Gini index is again twice the area between the Lorenz curve and the Line of Equality.

\begin{figure}[b]
	\centering
	\includegraphics[width = \textwidth]{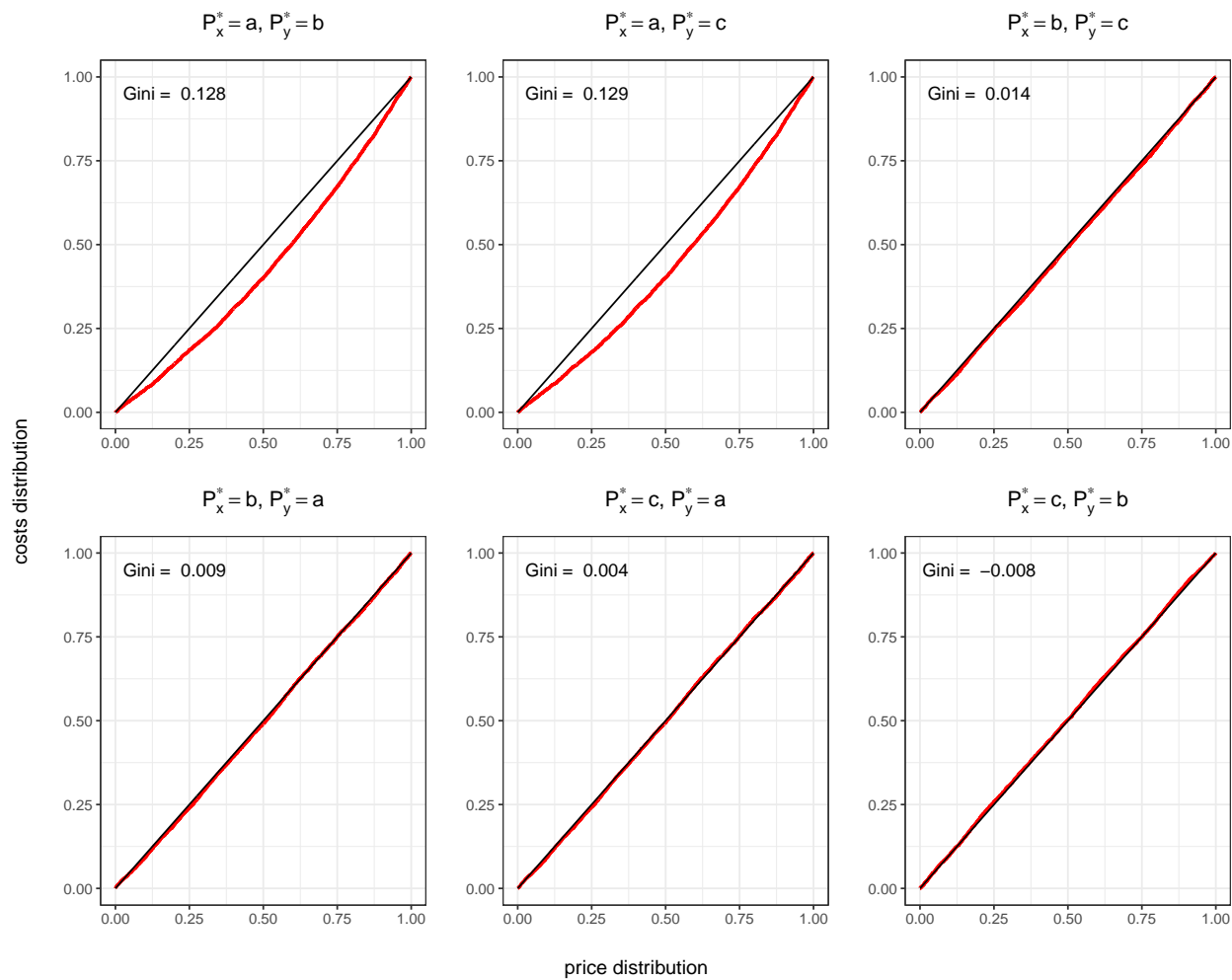}
	\caption{Ordered Lorenz curves (red) and Line of Equality (black) for the different tariff plans and costs over a two year horizon. An ordered Lorenz curve below the Line of Equality indicates that tariff plan $y$ is preferred over $x$.}
	\label{fig: ordered Lorenz}
\end{figure}

Figure~\ref{fig: ordered Lorenz} shows a two-way comparison of the three tariff plans resulting in six ordered Lorenz curves.
As shown in the top left panel for tariff plan $a$, the contracts with lowest $R_i$ are priced too high with respect to their costs (i.e., contribute disproportional to the total price than to the total costs), and vice versa for the contracts with high~$R_i$. 
This results in a convex curve below the Line of Equality. 
An ordered Lorenz curve corresponding to the Line of Equality (for instance in the bottom panels) indicates that the cumulative percentage of the costs equals the cumulative percentage of the prices, both in increasing relativities. 
Consequently, the alternative tariff $y$ is not preferred over the current tariff $x$.
Tariff plan $y$ is preferred over $x$ in case the ordered Lorenz curve is further away from the Line of Equality, indicating that more contracts have a relativity further away from 1.
The more convex the ordered Lorenz curve, the more tariff $y$ can recognize deficiencies in tariff $x$.
The latter is equivalent with a larger Gini index.
In this sense tariff plans $b$ and $c$ both clearly dominate scheme $a$.
The advantage of tariff $c$ over $b$ is, however, minor. 

\iffalse
\citet{frees2014insurance} state that half of the Gini index for ordered Lorenz curves can be interpreted as an average profit under the \colg{current tariff plan $P^\star_x$}. 
This average is under the implicit assumption that only contracts are retained in the portfolio with a relativity smaller than $R_i$ for an arbitrary $i$, average is taking over all possible $i$.
The maximal profit would be obtained when only those contracts are retained with relativity $R_i \leq R^\star$, where $R^\star$ is associated with the point on the Lorenz curve where the tangent is parallel to the Line of Equality.
\colg{Therefore, even if the service provider opts to keep a tariff plan $P^\star_x$ that differentiates less, the better differentiating tariff $P^\star_y$ plan still provides insight.
It gives the service provider insight which machines would be interesting to have in the contract portfolio and which machines are better avoided.
Using the insight of tariffs $b$ and $c$ while keeping tariff $a$ could lead a profit of $6\%$ on average.}
\fi

\FloatBarrier
\section{Conclusions}
\iffalse
As more manufacturers shift their focus from selling products to end solutions, full-service maintenance contracts gain traction in the business world. 
These contracts cover all maintenance related costs during a predetermined horizon in exchange for a fixed service fee and relieve customers from uncertain maintenance costs. 
To guarantee profitability, the service fees should at least cover the expected costs during the contract horizon. 
As these expected costs may depend on several machine-dependent characteristics (e.g., operational environment), the service fees should also be differentiated based on the risk profile.
If not, customers that are less prone to high maintenance costs will not buy into or renege on the contract.
The latter can lead to adverse selection and leave the service provider with a maintenance-heavy portfolio, which may be detrimental to the profitability of the service contracts.
We contribute to the literature with a data-driven tariff plan based on the calibration of predictive models that take into account the different machine profiles. 
This conveys to the service provider which machine profiles should be attracted at which price. 
We demonstrate the advantage of a differentiated tariff plan and show how it better protects against adverse selection.
\fi
%We \colb{show how to find the subset of contracts that maximize the profit under a non-differentiated tariff plan making use of the insights provided by the differentiated tariff plan.}
We study full-service maintenance agreements that cover all maintenance costs over a predetermined time horizon in exchange for a fixed premium. 
The stochastic and machine-dependent nature of the maintenance costs renders the determination of the premium challenging for the service provider. 
Inspired by insurance pricing, we use predictive analytics to determine the break-even price based on customer- and machine-dependent characteristics.
We build a simulation engine to simulate data that reflect this heterogeneity in maintenance occurrences and their costs and we show the economic value of price differentiation.

We demonstrate how price differentiation can help withstand adverse selection in case of a machine population with diverse maintenance costs.
%The latter would be detrimental for the profitability of the contract business of the service provider.
We provide visual tools, such as quantile plots and (ordered) Lorenz curves, and  quantitative measures, i.e. the Gini coefficient, to provide insight in the performance of the resulting tariff plans with respect to the heterogeneity of the machine population and with respect to each other.

We propose a couple of directions for further research, some of which are relaxations of the assumptions made in this paper.
First, we assumed the independence between occurrence, or frequency, of failures and their associated costs. 
The insurance pricing literature proposes to deal with dependence between occurrence and costs of claims by making use of for instance copulas. 
This approach could also be transferred to pricing full-service maintenance contracts. 
Second, we have demonstrated our approach on simulated data. 
Although our simulation engine reflects a lot of properties of real-world data, a validation of our approach on a real dataset seems a logical next step.
Third, our method relies on observable and measurable covariates to classify machines.
%This approach also implies that the service provider should keep track of the necessary risk factors in their contract databases in a similar fashion as insurance companies do. 
We assume that most of the heterogeneity between machines is captured by these measurable covariates or risk factors.
In the event that some heterogeneity is not explained by the observed covariates, we could rely on frailties, also known as random effects, to model the unobserved heterogeneity.

\FloatBarrier
\section*{Acknowledgments}
The authors thank the anonymous referees and the editor for useful comments which led to significant improvements of the paper.
\bibliography{bib}
\bibliographystyle{apalike}
\FloatBarrier
\newpage
\appendix
\section{List Of Notations}\label{appendix: notation}
\printnomenclature[0.9in]
\nomenclature[01]{$t_{0,i}$}{start of the contract for machine $i$}
\nomenclature[02]{$\Delta t$}{duration of the contract}
\nomenclature[03]{$n_m$}{number of preventive maintenance actions during contract}
\nomenclature[04]{$S_j$}{costs of preventive maintenance $j$}
\nomenclature[05]{$n_f$}{number of different failure types}
\nomenclature[05]{$n_{f,\text{minor}}$}{number of different minor failure types}
\nomenclature[06]{$N_{f,i}$}{number of failures of type $i$}
\nomenclature[07]{$X_{i,k}$}{costs of the $k$th failure of type $i$}
\nomenclature[08]{$C(\Delta t)$}{total costs covered during contract of length $\Delta t$}
\nomenclature[09]{$F(\Delta t)$}{total failure costs covered during contract of length $\Delta t$}
\nomenclature[10]{$M(\Delta t)$}{total preventive maintenance costs covered during contract of length $\Delta t$}
\nomenclature[11]{$P$}{price of the contract}
\nomenclature[12]{$P^\star$}{break-even or technical price of the contract}
\nomenclature[12]{$\Delta t_{\text{PM}}$}{preventive maintenance interval}
\nomenclature[12]{$S_T(t\vert t_0)$}{survival function for the time-to-next-failure}
\nomenclature[12]{$\lambda_\cdot (t)$}{failure intensity function}
\nomenclature[13]{$\bm{x}_1$,$\bm{\chi}_1$}{vector of fixed risk factors}
\nomenclature[15]{$\bm{w}$,$\bm{w}_y$,$\bm{w}_c$,$\bm{z}$}{vector of risk factors}
\nomenclature[14]{$\bm{x}_2(t)$,$\bm{\chi}_2(t)$}{vector of time-dependent risk factors}
\nomenclature[16]{$y$}{failure type}
\nomenclature[17]{$m$}{preventive maintenance}
\nomenclature[18]{$f_m$}{minor failure}
\nomenclature[20]{$f_c$}{catastrophic failure}

\section{Pseudo-code of the simulation engine}
\label{appendix: algo}
\FloatBarrier
\begin{algorithm}[H]
\caption{Simulation of a realistic dataset of maintenance and failure events}
\label{alg: simulation}
\begin{algorithmic}[1]
\State initialization;
\State timeLinesPortfolio $ =\emptyset$;
\For{$i\ \textbf{in}\ \{1,...,n\}$}
    \State $t = 0$;
	\State $t_{\text{previous}} = 0$;
	\State timeLineMachine $ =\emptyset$;
	\While{$t \leq t_{\text{obs},i}$}
        \State $t$ = getFailureTime($\bm{x}_i$, $t_{\text{previous}}$);
        \State $t_{\text{previous}} = t$;
        \State $y$ = getFailureType($\bm{x}_i$, $t$);
		\State $x$ = getFailureCosts($\bm{x}_i,\ y$);
		\State timeLineMachine $\leftarrow$ $[i,t,y,x]$;
    \EndWhile
    \State timeLineMachine $\leftarrow [i,t_{\text{obs},i},\text{end},0]$;
	\State $n_\text{PM} = \left \lfloor \frac{t_{\text{obs},i}}{\Delta t_{\text{PM}}} \right\rfloor$;
	\For{$j\ \textbf{in}\ \{1,...,n_\text{PM}\}$}
		\State $s$ = getMaintenanceCosts($\bm{x}_i$);
		\State timeLineMachine $\leftarrow [i,j\cdot \Delta t_{\text{PM}},\text{PM},s]$;
	\EndFor
	\State timeLinesPortfolio $\leftarrow$ timeLineMachine;
\EndFor
\end{algorithmic}
\end{algorithm}

\FloatBarrier
\section{Results of the calibration of the predictive models in our numerical experiment}
\label{appendix: param en est}
We present the calibrated parameters and their $95\%$-confidence intervals of the predictive models for the simulation settings as introduced in Section \ref{simulation} in Table \ref{tab: pricing}.
These are obtained by the calibration of a single simulation run.
\begin{landscape}
\begin{table}[tb]
\caption{Overview of simulation parameters, estimated values and confidence intervals for the calibration and the pricing schemes. As different covariates are taken into account, the outcome of the calibration will be different for each pricing scheme.}
\label{tab: pricing}
\centering
\begin{small}
\begin{tabular}{*4c}
\toprule
Parameter & Simulation &\multicolumn{2}{c}{Calibration}\\
& & estimate & $95\%$ conf.  \\
 &  &  &  interval\\
\midrule
$\alpha_{\lambda_0}$ & $0.5$ & $0.472$ & $(0.443,0.502)$\\
$t_\text{PM}$ & $1$ & - & -\\
$\kappa_{\lambda_0}$ & $0.623$& $0.599$&$(0.571,0.626)$\\
$\gamma_{\lambda_0}$& $0.1$&$0.110$ & $(0.097,0.124)$\\
$\bm{\beta}_1$ &$-0.2$ & $-0.186$ & $(-0.214,-0.157)$\\
& $0.3$ & $0.311$ & $(.282,.339)$\\
& $0.4$ & $0.411$ & $(0.382,0.439)$\\
& $-0.1$ & $-0.097$ & $(-0.125,-0.068)$\\
$\bm{\beta}_2$&$0.1$&$0.101$ & $(0.066,0.136)$\\
$\alpha_{c}$ & $0.2$ & $0.198$& $(0.195,0.202)$\\
$\kappa_{c}$ & $2$& $1.990$&$(1.936,2.045)$\\
$\bm{\alpha}_1$ & $0.9$ & $0.948$ & $(0.845,1.052)$\\
& $0.4$ & $0.363$ & $(0.268,0.459)$\\
& $0.1$ & $0.055$ & $(-0.038,0.149)$ \\
& $0.0$ & $0.027$ & $(-0.066,0.121)$\\
& $0.1$ & $0.141$ & $(0.048,0.235)$\\
$\bm{\alpha}_2$ & $0.9$ & $0.956$ & $(0.853,1.058)$\\
& $0.5$ & $0.474$ & $(0.380,0.568)$\\
& $0.0$ & $-0.033$ & $(-0.125,0.059)$\\
& $0.2$ & $0.213$ & $(0.121,0.306)$\\
& $0.2$ & $0.232$ & $(0.140,0.324)$\\
$\bm{\gamma}_m$ & $20$ & $20.154$ & $(19.798,20.509)$\\
& $3$ & $2.984$ & $(2.931,3.038)$\\
$\bm{\gamma}_1$&$20$ & $20.158$ & $(19.530,20.785)$\\
& $3$ & $2.984$ & $(2.893,3.081)$\\
$\bm{\gamma}_2$&$30$ & $29.474$ & $(28.626,30.322)$\\
& $5$ & $5.100$& $(4.957,5.253)$\\
$\theta$ & $0.5$ & $0.509$ & $(0.264,0.753)$ \\
$\bm{\gamma}_c$ & $20$ & $19.813$  & $(18.909,20.718)$ \\
& $10$ & $10.070$& $( 9.625,10.558)$ \\
\bottomrule
\end{tabular}%
\begin{tabular}{*6c}
\toprule
\multicolumn{2}{c}{Pricing scheme $a$} & \multicolumn{2}{c}{Pricing scheme $b$}& \multicolumn{2}{c}{Pricing scheme $c$}\\
 estimate & $95\%$ conf. & estimate & $95\%$ conf. & estimate & $95\%$ conf.\\
  & interval &  & interval &  & interval\\
\midrule
$0.620$ & $(0.587,0.654)$ & $0.482$ & $(0.453,0.512)$&  $0.472$ & $(0.443,0.502)$\\
- & - & - & - & - & -\\
$0.590$ & $(0.562,0.617)$& $0.590$ & $(0.563,0.618)$ & $0.599$ & $(0.571,0.626)$\\
$0.139$ & $(0.123,0.157)$& $0.108$ & $(0.095,0.122)$ &  $0.110$ & $(0.097,0.124)$\\
- & - & $-0.196$ & $(-0.224,-0.168)$& $-0.186$ & $(-0.214,-0.157)$\\
- & - & $0.316$ & $(0.287,0.344)$ & $0.311$ & $(.282,.339)$\\
- & - & $0.415$ & $(0.387,0.444)$ & $0.411$ & $(0.382,0.439)$\\
- & - & $-0.103$ & $(-0.131,-0.074)$ & $-0.097$ & $(-0.125,-0.068)$\\
- & - & - & - & $0.101$ & $(0.066,0.136)$\\
$0.198$ & $(0.195,0.202)$& $0.198$ & $(0.195,0.202)$ & $0.198$& $(0.195,0.202)$\\
$1.990$ & $(1.936,2.045)$ & $1.990$& $(1.936,2.045)$ &$1.990$&$(1.936,2.045)$\\
$1.206$ & $(1.160,1.252)$ & $0.948$ & $(0.845,1.052)$ &$0.948$ & $(0.845,1.052)$\\
- & - & $0.363$ & $(0.268,0.459)$ & $0.363$ & $(0.268,0.459)$\\
- & - & $0.055$ & $(-0.038,0.149)$ & $0.055$ & $(-0.038,0.149)$\\
- & - & $0.027$ & $(-0.066,0.121)$ &$0.027$ & $(-0.066,0.121)$\\
- & - & $0.141$ & $(0.048,0.235)$ & $0.141$ & $(0.048,0.235)$\\
$1.368$ & $(1.322,1.413)$& $0.956$ & $(0.853,1.058)$ & $0.956$ & $(0.853,1.058)$\\
- & - & $0.474$ & $(0.380,0.568)$ & $0.474$ & $(0.380,0.568)$\\
- & - & $-0.033$ & $(-0.125,0.059)$ & $-0.033$ & $(-0.125,0.059)$\\
- & - & $0.213$ & $(0.121,0.306)$ & $0.213$ & $(0.121,0.306)$\\
- & - & $0.232$ & $(0.140,0.324)$ & $0.232$ & $(0.140,0.324)$\\
$20.154$ & $(19.798,20.509)$ & $20.154$ & $(19.798,20.509)$ & $20.154$ & $(19.798,20.509)$\\ 
$2.984$ & $(2.931,3.038)$ & $2.984$ & $(2.931,3.038)$ & $2.984$ & $(2.931,3.038)$\\
$20.158$ & $(19.530,20.785)$ & $20.158$ & $(19.530,20.785)$ & $20.158$ & $(19.530,20.785)$\\
$2.984$ & $(2.893,3.081)$ & $2.984$ & $(2.893,3.081)$ & $2.984$ & $(2.893,3.081)$\\
$29.474$ & $(28.626,30.322)$ & $29.474$ & $(28.626,30.322)$ & $29.474$ & $(28.626,30.322)$\\
$5.100$& $(4.957,5.253)$ & $5.100$& $(4.957,5.253)$ & $5.100$& $(4.957,5.253)$\\
$0.509$ & $(0.264,0.753)$ & $0.509$ & $(0.264,0.753)$ & $0.509$ & $(0.264,0.753)$ \\
$19.813$  & $(18.909,20.718)$ & $19.813$  & $(18.909,20.718)$ & $19.813$  & $(18.909,20.718)$\\ 
$10.070$& $( 9.625,10.558)$ & $10.070$& $( 9.625,10.558)$ & $10.070$& $( 9.625,10.558)$\\
\bottomrule
\end{tabular}
\end{small}
\end{table}
\end{landscape}
\FloatBarrier

\FloatBarrier
\iffalse
\section{Double lift chart}
\label{ap: double lift}
A double lift chart makes the direct comparison between two tariff plans $P^\star_x$ and $P^\star_y$ possible.
This chart is obtained by sorting the contracts in ascending relativity $R_i$.
Next, we divide the contracts into equal bins.
Finally, we calculate the average percentage error $\frac{P}{C}-1$ for both tariffs where $P$ is the average price and $C$ the average costs in the bin.
The better tariff plan will have percentage errors closer to zero, indicating that those premiums match the actual costs better.
Figure~\ref{fig double lift} displays the lift charts comparing the different tariff plans. 
It is clear that tariff plans $b$ and $c$ outperform plan $a$, since the percentage errors are a lot smaller.
The difference between tariffs $b$ and $c$ is less obvious however tariff $c$ outperforms $b$ slightly. 
\begin{figure}[h]
    \centering
    \begin{tikzpicture}
	\definecolor{lightb}{RGB}{153,204,255}
	\draw [fill = black, black] (4.6,-0.05) rectangle (4.7,0.05) {};
	\node [right] at (4.7,0.01) {$P^\star_x$};
	\draw [fill = red, red] (5.6,-0.05) rectangle (5.7,0.05) {};
	\node [right] at (5.7,0.01) {$P^\star_y$};
\end{tikzpicture}
	\centering
	\includegraphics[width = \textwidth]{figures/liftchart.eps}
	\caption{Lift chart over a two year horizon comparing the different tariff plans. Clearly tariff plans $b$ and $c$ outperform plan $a$.}
	\label{fig double lift}
\end{figure}
\fi
\FloatBarrier
\newpage
\section{Simulation study to verify the calibration method}
\label{appendix verification}
\begin{figure}
	\centering
	\includegraphics[width = \textwidth]{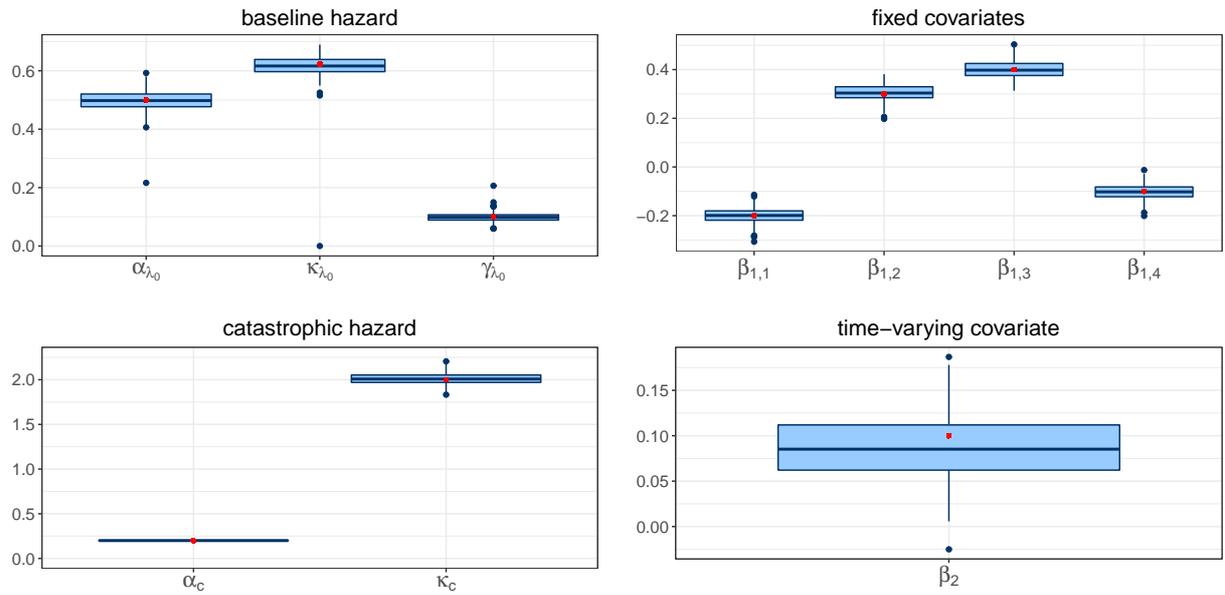}
	\caption{Maximum likelihood estimates for the time-to-failure parameters (red dot: simulation parameter values)}
	\label{fig: MLE TTF}
\end{figure}

\begin{figure}
	\centering
	\includegraphics[width = \textwidth]{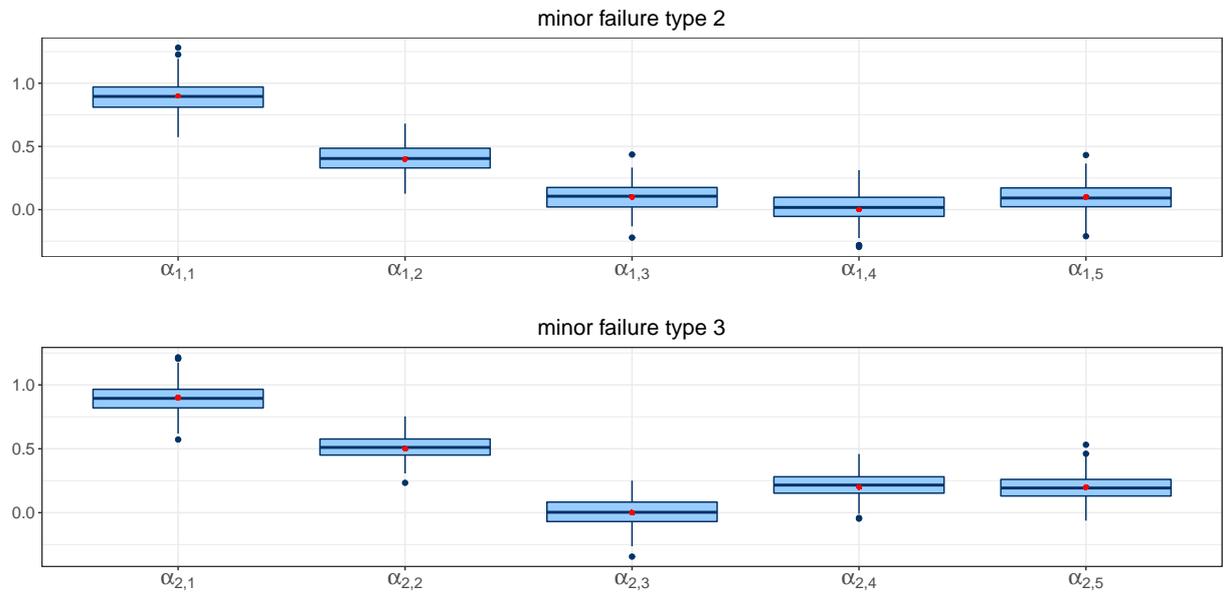}
	\caption{Maximum likelihood estimates for the multinomial distribution parameters (red dot: simulation parameter values)}
	\label{fig: multinom}
\end{figure}

\begin{figure}
	\centering
	\includegraphics[width = \textwidth]{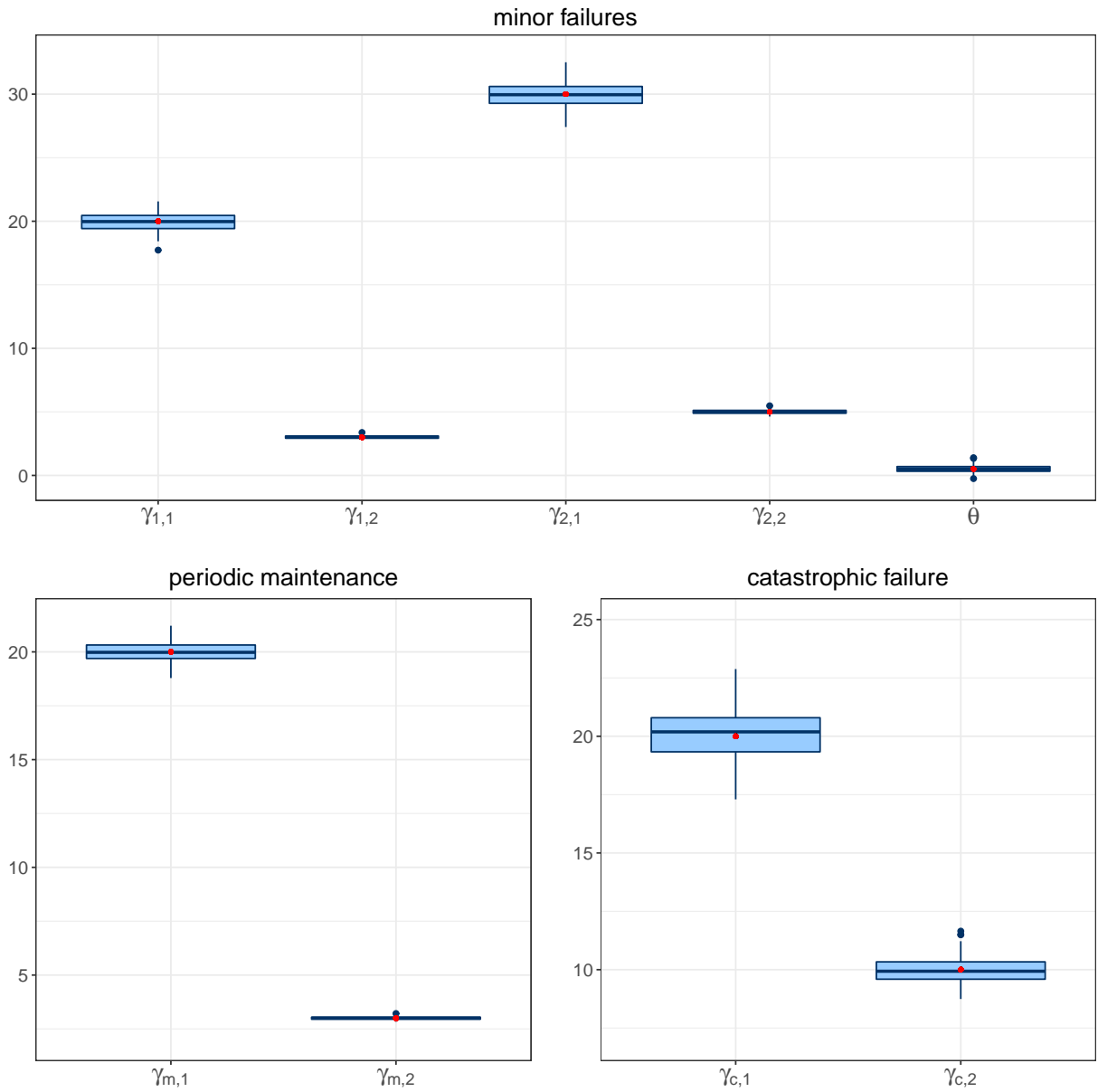}
	\caption{Maximum likelihood estimates for the costs distribution parameters (red dot: simulation parameter values)}
	\label{fig: costs}
\end{figure}

To verify the calibration of our predictive models, we simulate $250$ replications of the simulation setting described in Section \ref{simulation}.
The number of machines for this verification is equal to $n = 1000$, in contrast to $n = 5000$ in Section \ref{simulation}, that we used to illustrate our pricing methodology.
Each replication leads a single estimate for the simulation parameters.
Consequently it is possible to compare these estimates with the simulation parameters by means of a boxplot.
The boxplots for the time-to-failure model parameters, multinomial distribution parameters and the costs distribution parameters can be found on Figures \ref{fig: MLE TTF}, \ref{fig: multinom} and \ref{fig: costs} respectively and the averages of the estimates in Table \ref{tab: test sim param}.
The average estimates are very close to the simulation parameters and this confirms the accuracy of calibration strategy.
\begin{table}[h]
\caption{Simulation parameters and average estimates}
\label{tab: test sim param}
\centering
\begin{tabular}{ccc}
\toprule
parameter & simulation   & average\\
 &  value  & estimate\\
\midrule
$\alpha_{\lambda_0}$ & $0.5$ & $0.496$\\
$\kappa_{\lambda_0}$ & $0.623$&$0.613$\\
$\gamma_{\lambda_0}$& $0.1$&$0.099$\\
$\bm{\beta}_1$ &$(-0.2,0.3,0.4,-0.1)$ &$(-0.199,0.305,0.399,-0.101)$\\
$\bm{\beta}_2$&$0.1$&$0.088$\\
$\alpha_{c}$ & $0.2$ &$0.200$\\
$\kappa_{c}$ & $2$&$2.009$\\
$\bm{\alpha}_1$ & $(0.9,0.4,0.1,0.0,0.1)$ & $(0.894,0.409,0.101,0.017,0.095)$\\
$\bm{\alpha}_2$ & $(0.9,0.5,0.0,0.2,0.2)$ & $(0.892,0.514,.001,0.215,0.195)$\\
$\bm{\gamma}_m$ & $(20,3)$ & $(20.005,3.001)$\\ 
$\bm{\gamma}_1$&$(20,3)$ & $(19.945,3.013)$\\
$\bm{\gamma}_2$&$(30,5)$ & $(29.986,5.008)$\\
$\theta$ & $0.5$ & $0.511$\\
$\bm{\gamma}_c$ & $(20,10)$& $(20.092,9.976)$\\ 
\bottomrule
\end{tabular}
\end{table}

\end{document}